Review

# Multi-Channel Man-in-the-Middle Attacks Against Protected Wi-Fi Networks: A State of the Art Review

**_Manesh Thankappan[a, *], Helena Rifà-Pous[a,b], Carles Garrigues[a,b]_**

[a]*Estudis d'Informàtica Multimèdia i Telecomunicació,Internet Interdisciplinary Institute (IN3), Universitat Oberta de Catalunya (UOC), Barcelona, Spain*

[b]*Center for Cybersecurity Research of Catalonia (CYBERCAT), Barcelona, Spain*

A B S T R A C T

Multi-Channel Man-in-the-Middle (MitM) attacks are special MitM attacks capable of manipulating encrypted wireless frames between two legitimate endpoints. Since its inception in 2014, attackers have been targeting Wi-Fi networks to perform different attacks, such as cipher downgrades, denial of service, key reinstallation attacks (KRACK) in 2017, and recently FragAttacks in 2021, which widely impacted millions of Wi-Fi devices, especially IoT devices. To the best of our knowledge, there are no studies in the literature that holistically review the different types of Multi-Channel MitM enabled attacks and analyze their potential impact. To this end, we evaluate the capabilities of Multi-Channel MitM and review every reported attack in the state of the art. We examine practical issues that hamper the total adoption of protection mechanisms, i.e., security patches and Protected Management Frames (PMF), and review available defense mechanisms in confronting the Multi-Channel MitM enabled attacks in the IoT context. Finally, we highlight the potential research problems and identify future research lines in this field.

## 1. Introduction

WLANs are broadly employable in several networking applications because of their flexibility, mobility, and availability. With the influx of the Internet of Things (IoT), Wi-Fi devices operating on the 802.11 standards are now gaining widespread deployment everywhere. Unfortunately, WLANs are susceptible to a broad array of wireless security attacks. Man-in-the-middle (MitM) attacks are a common form of security attack towards wireless networks that allow attackers to catch and manipulate communication between two end devices. One of the advanced MitM attacks is the Multi-Channel MitM (MC-MitM) attack that can manipulate the encrypted network traffic, as presented in (Vanhoef & Piessens, 2014). Since (Vanhoef & Piessens, 2014), MC-MitM attacks have been a trend in exploiting various Wi-Fi Protected Access (WPA) protocols in personal and enterprise networks. These kinds of attacks include denial of service (DoS), security downgrades, key reinstallations, and other vendor-specific exploits. The MC- MitM attack makes use of two different channels that facilitates the attacker to forward frames between both channels so that he can legitimately manipulate (e.g., block, delay, modify, inject, replay) encrypted frames between clients and the access point (AP) in a WLAN.

The Wi-Fi Alliance (WFA) and leading device manufacturers started noticing the MC-MitM attacks after the disclosure of a massive key reinstallation vulnerability (CVE-2017-13077) in the mid of 2017 (Vanhoef & Piessens, 2017). This was the first non-vendor specific vulnerability (as it is found in 802.11 standards) that could be exploited by MC-MitM enabled attack known as key reinstallation attack (KRACK), which abuses severe vulnerabilities, such as nonce and replay counter reuse during 4-way handshake mechanisms in the WPA and WPA2 certified devices. This vulnerability makes MC-MitM attackers more effective as they can trivially decrypt Wi-Fi frames, especially from Linux and Android devices. To resolve key reinstallation vulnerabilities, the Wi-Fi Alliance and some affected Wi-Fi chip manufacturers released patches. Available patches are only applicable to powerful Wi-Fi clients (e.g., laptops, smartphones, routers, etc.). However, many Wi-Fi devices cannot be patched because some companies do not provide them, especially IoT devices suffer from this issue. Some constraints like low computing capacities or specific network settings also impede the adoption of patches on IoT devices. This situation pushes millions of WPA or WPA2 devices, especially IoT devices, to remain vulnerable to MC-MitM attackers. In 2019, a security research company tested several commercially available Wi-Fi devices and reported that 90% of them are vulnerable to key reinstallation attacks (Security Focus, 2019). Another recent exploratory study (Aug 2020) presented in (Freudenreich et al., 2020) critically indicates that overall, 65% of Wi-Fi and IoT devices tested are vulnerable to key reinstallation attacks. Regarding WPA3, although it provides improved security features in terms of encryption, it does not prevent KRACK on its own (Vanhoef, 2017b). This is because WPA3 executes the same 4-way handshake mechanism that is vulnerable in the same way as the one present in WPA and WPA2 protocols. The resilience of WPA3 devices towards KRACK also solely depends on the application of patches during the WPA3 certification process (Krischer, 2019).

* *Corresponding author.*
E-mail address: mthankappan@uoc.edu (**Manesh Thankappan**), hrifa@uoc.edu (**Helena Rifà-Pous**), cgarrigueso@uoc.edu (**Carles Garrigues**).

In mid of May 2021, Vanhoef presented a set of new MC-MitM enabled attacks dubbed as FragAttacks (fragmentation and aggregation attacks) (Vanhoef, 2021a), which abuse serious security vulnerabilities (CVE-2020-24586,87,88) due to the lack of proper authentication in the aggregation and fragmentation features of 802.11 standards. This is another non-vendor specific vulnerability affecting every Wi-Fi device, including the new WPA3. FragAttacks enable attackers to inject packets into protected Wi-Fi networks and then capture client´s sensitive data. WFA has released patches for these vulnerabilities, and other leading device vendors are currently releasing patches. The arrival of FragAttacks brings big concern in terms of securing IoT devices as such devices rarely receive patches and can experience the same difficulties as KRACK patching in upcoming years.

Generally, besides patches, another solution to counter various MitM or DoS attacks is the use of 802.11w standard or Protected Management Frames (PMF), which provides integrity protection for wireless frames (Philipp Ebbecke (Wi-Fi Alliance), 2020). However, many existing Wi-Fi clients in our home or office settings, especially IoT devices, may not always comply with PMF. A significant reason is that PMF was vendor-specific and was optional for currently available WPA or WAP2 devices. Only some Cisco devices provide client support for the PMF standard. A new survey on Wi-Fi security risks presented in (Reyes et al., 2020) critically points out that around 87% of analyzed routers do not comply with PMF standards.

The remarkable point is that MC-MitM attacks can easily circumvent PMF protection as attackers utilize certain pre-authenticated management frames which are not protected even when PMF is enabled. Once MC-MitM is acquired, attackers can plan for several attacks. For example, they can trigger FragAttacks or KRACK as the PMF standard itself is vulnerable to such attacks (CVE-2017-13081). Additionally, the MC-MitM attackers can exploit several inherent PMF vulnerabilities (e.g., channel switch attacks, jam genuine channel switch announcements, forge reassociation frames) more practically and eventually cause a potential deadlock or DoS on PMF-capable networks (Vanhoef et al., 2018). These attacks are hard to detect because the attacker requires merely a single forged frame for the impact.

Similarly, another pertinent issue is that, even though new WPA3 routers enter our domestic networks, they must be configured to operate in transition mode to accommodate many PMF incapable or legacy devices. In this case, MC-MitM attackers may target and hijack such devices connected to WPA3 routers and challenge their security. This situation may persist for several years because millions of WPA or WPA2 devices are currently deployed everywhere. However, it is not a good practice to close eyes from the risk of possible MC-MitM attacks.

Detecting MC-MitM attacks is challenging because the attacker acquires MitM position between an already connected client and AP in a WLAN without disconnecting clients from the legitimate network. Most importantly, the MC-MitM attacker uses a rogue AP that behaves as a normal AP in a WLAN. He neither floods the Wi-Fi medium with deauthentication frames nor performs any other dubious activities while acquiring the MitM position and tricks end devices to believe that they are communicating with each other directly. So to correctly differentiate between the normal and dubious activities, some prudent mechanisms are required. In mitigating MC-MitM attacks, some mechanisms have been proposed in the literature. Amongst such defense mechanisms, we perceive that operating channel validation (Vanhoef et al., 2018) and beacon protection (Vanhoef et al., 2020) mechanisms can considerably harden these attacks. However, these mechanisms still allow partial MitM attacks or block MC-MitM attacks if they originate from outsiders or unauthenticated users. The significant problem that persists is how to block potentially such malicious insider MC-MitM attacks, and this vulnerability remains open in all WPA standards, including WPA3.

Currently reported MC-MitM attacks so far impact WPA, WPA2, and WPA3 devices. FragAttacks are the most recent ones in the series of MC-MitM attacks. This shows that currently incorporated MC-MitM defense mechanisms in 802.11 standards are not yet really used in practice. Our analysis also revealed that most of the existing mechanisms are not flexible to implement in IoT environments because they mandate installing additional security modules, configuring their new solutions on home routers or every Wi-Fi client. We highlight the point that there are several IoT devices in a smart environment, and the defense mechanism cannot be based on the premise that all these devices will have to be modified, updated, or replaced by new ones. The technical overhead on ordinary people is also considerably high when deploying existing defense mechanisms due to complex configurations, setting up specific networks, firmware installation, etc. Traditional IDS like SNORT are also ineffective in confronting this kind of MitM attack. This is because SNORT works at the network layer and cannot detect MC-MitM attacks at the link layer.

The above issues shed light on the fact that preventing MC-MitM attacks can be difficult in practice, and especially if IoT devices have limited protections against them. Therefore, IoT environments need imperative developments against these attacks and are essential due to the increased influx of IoT devices to our smart environments.

**Contributions of the Paper.** The main contributions of the paper are:

1. An in-depth evaluation of MC-MitM attack´s capabilities in manipulating protected Wi-Fi communications, in particular, WPA, WPA2 and WPA3 networks, and examining whether attacks discovered for WPA2 are still possible in WPA3.
2. A thorough review and a classification of MC-MitM enabled attacks.
3. An analysis of possible security impacts of MC-MitM attacks.
4. An examination of challenges in adopting general protection mechanisms such as security patches and PMF against MC-MitM attacks.
5. A technical feasibility analysis of existing defense mechanisms for MC-MitM enabled attacks in IoT context.
6. An analysis of potential research problems, challenges and future research approaches.

**Organization.** The paper's remainder is organized as follows. Section 2 briefly outlines protected Wi-Fi networks and fundamentals of MitM attacks. In Section 3, detailed technical setup and inner workings and classifications, specialities of MC-MitM attacks are presented. Section 4 reviews recent MC-MitM enabled attacks, examines significant difficulties in adopting security patches and PMF against MC-MitM attacks, and analyses how MC-MitM attacks impact new WPA3 networks. In Section 5, the existing detection mechanisms for combating MC-MitM attacks are reviewed followed by their technical feasibility analysis in the IoT context. Section 6 discusses identified research problems, challenges, and future research approaches in this field. Finally, Section 7 concludes our research analysis.

## 2. Protected Wi-Fi networks and MitM attacks

In this section, we explore various security protocols of 802.11 standards, including the PMF used for protecting Wi-Fi communication, and highlight the related issues in terms of MitM attacks. We provide fundamentals of Wi-Fi based MitM attacks and evaluate how rogue AP MitM attackers manipulate protected Wi-Fi communication. In this paper, the term "manipulate" is used to represent the attacker's ability to reliably intercept and perform operations, such as exchanging, blocking, forging, modifying, replaying, injecting, or decrypting link-layer wireless traffic using the MitM position.

### *2.1. A security analysis of protected Wi-Fi networks*

IEEE 802.11i standard defines protected Wi-Fi networks with more robust security solutions to the 802.11 standard. IEEE 802.11i is also known as a Robust Security Network (RSN) (Frankel et al., 2007). To provide link-layer protection for Wi-Fi communication under 802.11i, the Wi-Fi Alliance maintains three security certifications, namely WPA (Wi-Fi Protected Access), WPA2 (Wi-Fi Protected Access II), and WPA3 (Wi-Fi Protected Access III). As encryption or data confidentiality protocol, Temporal Key Integrity Protocol (TKIP) is used in WPA and is optional in WPA2, while WPA2 and WPA3 protocols mandate Advanced Encryption Standard (AES). As a data integrity protocol, WPA uses a Message Integrity Check (MIC) known as Michael algorithm (Beck & Tews, 2009), WPA2 and WPA3 mandate Counter Mode CBC-MAC Protocol (CCMP) for their personal networks, and the new Galois Counter Mode Protocol (GCMP) enhances WPA3 security for its enterprise networks (He & Mitchell, 2004; Vanhoef, 2017b). In this paper, instead of certifications, we may use terms such as devices, or networks interchangeably depending on context.

### *2.2. Connection establishment in WPA, WPA2 and WPA3 networks*

According to 802.11i, when a client connects to a router or AP in a Basic Service Set (BSS) or WLAN, it passes through four phases: (i) network discovery, (ii) authentication, (iii) association, and (iv) 4-way handshake mechanism, as illustrated in Fig.1. This connection establishment is also known as 802.11 State Machine (Frankel et al., 2007). Following, we briefly discuss the four phases.

#### *2.2.1 Network discovery*

In a WLAN, APs show their network presence by periodically broadcasting beacons. A beacon includes the SSID (Service Set Identifier) or network name, MAC address, channel, and other capabilities of the AP. Next, the client device scans and lists available networks so that the user can select the appropriate network and manually enter the Wi-Fi passphrase or pre-shared key (PSK) that was already configured in the AP. Important to note that this passphrase is stored or cached in the Wi-Fi chip of the client and is never transmitted or exchanged in any of the frames. With the selected SSID, the client sends a probe request frame to verify whether a specific network is available or not, and this activity begins with the state machine. In response to this, the AP sends a probe response frame by acknowledging the availability of SSID. This finishes the network discovery. The steps of this phase are common for WPA, WPA2, and WPA3.

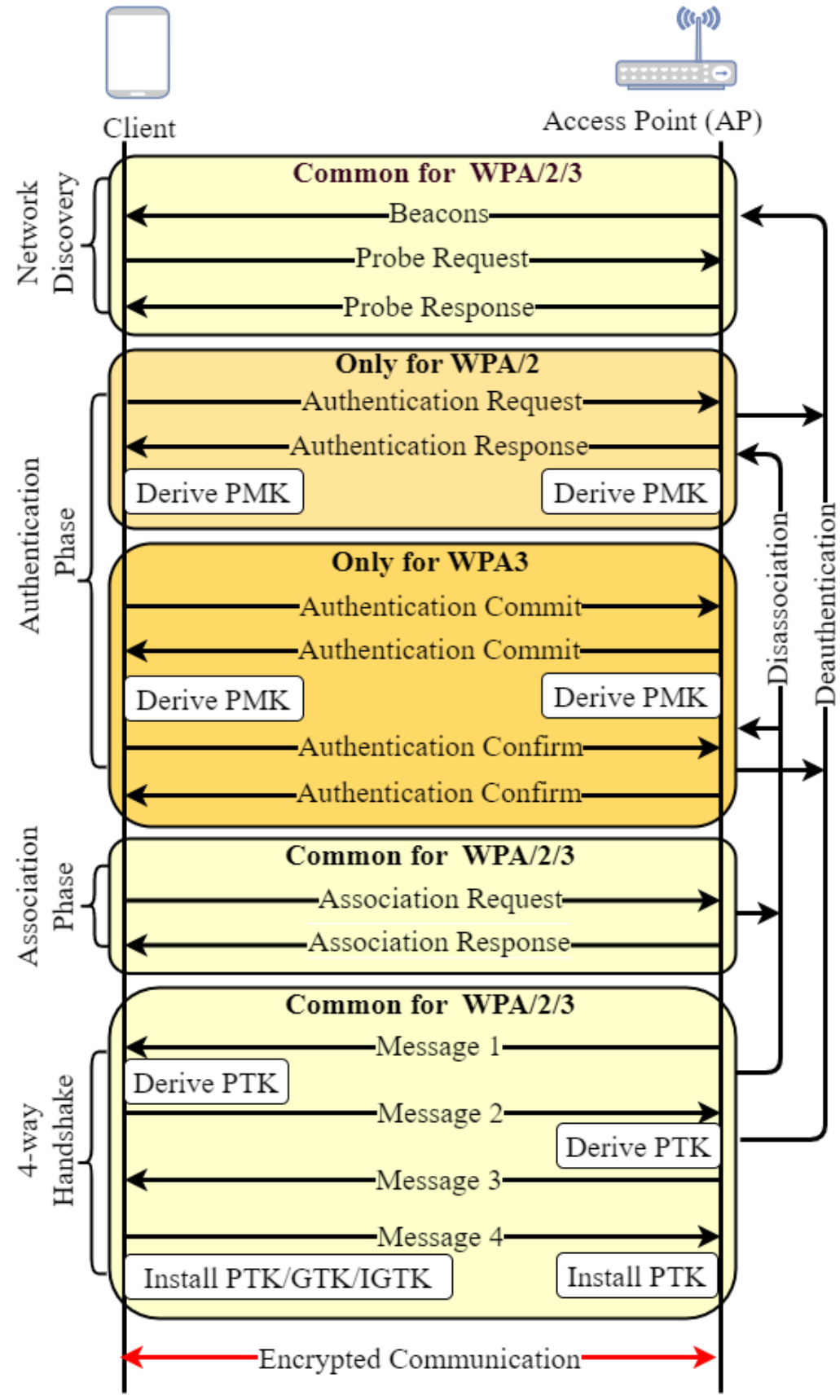

**Fig. 1. Generalized connection establishment between the AP and client** (Kohlios & Hayajneh, 2018; Vanhoef, 2017b)**.**

#### *2.2.2 Authentication*

During the authentication, the AP verifies the client's identity (MAC address) and registers it in its cache. As shown in Figure 1, the authentication phase has different steps according to the version of the security protocol. In WPA or WPA2, the client and AP exchange open authentication requests and response frames. Upon a successful authentication, a Pre-Master Key (PMK) is derived from the PSK on either side. On the other hand, the WPA3 protocol executes a new Dragonfly handshake (termed as Simultaneous Authentication of Equals, a.k.a SAE) by exchanging four authentication frames. Before this, both the client and AP generate a secret element known as Password Element (PWE) and two secret values. During the first two authentication frames (commit messages), the client and AP negotiate a PMK through Elliptic Curve

Diffie–Hellman (ECDH) key exchange technique (Kohlios & Hayajneh, 2018; Vanhoef & Ronen, 2020) . In the last two authentication messages, they confirm that both negotiated the same PMK. This way, the PMK is calculated using respective security protocol and cached on either device, maintaining the state machine. The generated PMK will be utilized in the 4-way handshake. After authentication, deauthentication frames from either client or AP will cause a disconnection from the network.

#### *2.2.3 Association*

As soon as the authentication ends, the client prepares to associates with the AP by forwarding an association request to negotiate required cipher suites such as TKIP/CCMP/GCMP. During the client's association, the AP keeps an association ID and sends an association response back. A client can be authenticated to many networks but can be associated with only one network at a time (Frankel et al., 2007). Note that with cached PMK, a client will be allowed to rejoin an already associated AP even after leaving the network, or a client can be quickly reconnected to the AP after an intermittent disconnection. Thus, the user or client does not need to enter a Wi-Fi passphrase again since the AP maintains its state machine or security association. This procedure is the same in WPA, WPA2, and WPA3. Like deauthentication, disassociation can occur at this stage, disconnecting the client. Finally, the 4-way handshake starts upon a successful association.

#### *2.2.4 4-Way handshake mechanism*

The 4-way handshake mechanism is the same in WPA, WPA2, and WPA3 protocols, and involves exchanging 4 EAPOL (Extensible Authentication Protocol over LAN). During this handshake, the AP and client derive a Pairwise Transient Key (PTK), also known as a session key, which is then used for encrypting the actual communication between them. To derive PTK, the PMK is used with other parameters which are: AMAC represents AP´s MAC address, CMAC represents client´s MAC address; AN represents AP´s random number, CN represents client´s random number; RC is the replay counter; PRF indicates Pseudo-Random Function. Finally, MIC (x, x, etc.) brings the Message Integrity Code created for the contents within the parentheses with derived PTK so that the AP or Wi-Fi client can verify whether this message is corrupted or not (He & Mitchell, 2004, Hiertz et al., 2010). Group-Transient Key (GTK) is independently derived at every AP and is the same for all the clients connected to it. Similarly, Integrity Group-Transient Key (IGTK) will be derived if PMF is enabled (see Section 2.3). Corresponding 4-way handshake message exchanges are summarized as follows.

- **Message 1: AP → Client**
  The AP sends [AMAC address, AN, and RC] to the client. With these values, the client derives PTK, i.e., PTK ← PRF (PMK, AN, SN, AMAC, CMAC).
- **Message 2**: **AP ← Client**
  Once PTK is derived, the client sends [CMAC, SN, RC, and MIC (CMAC, SN, RC)] to the AP.
- **Message 3: AP → Client**
  Once message 2 is received, AP verifies MIC and derives PTK. The AP also derives GTK (Group Transient Key) and then the AP sends back [AMAC, AN, RC+1, GTK and MIC (CMAC, SN, RC+1, GTK)] to the client.
- **Message 4: AP← Client**
  Once message 3 is received, the client sends [CMAC, SN, RC+1, and MIC (CMAC, SN, RC+1)] to the AP to acknowledge reception of message 3 successfully. Consequently, both the AP and client will install PTK and GTK.

With the 4-way handshake, both the AP and client complete the state machine and stay connected. During this phase, deauthentication or disassociation can happen due to various reasons. Once end devices install security keys, the pairwise data communication between the AP and client will be encrypted (at the link layer) by the session key PTK using negotiated ciphers. The AP uses GTK to encrypt broadcast or multicast frames to communicate with every associated client.

As far as WPA and WPA2 are concerned, the foremost issue is that they are vulnerable to brute force or dictionary attacks, which aid attackers in retrieving security keys and potentially decrypt previously encrypted sessions. This happens because the generated PMK is the same for all clients. However, WPA3 solves this issue prominently by using a Dragonfly handshake that not only increases the entropy of the PMK but also ensures robust authentication/key exchange through Elliptic Curve Cryptography (ECC) and strong encryption through AES-GCMP. Therefore, offline dictionary attacks and the compromise of previous sessions (forward secrecy) are prevented since the derived PMK is independent of the PSK, and each client has a different PMK.

On the other hand, although data frames (actual communication between end devices) are protected using security protocols, all the management frames during the network discovery, authentication, and association phases are left unprotected as they are exchanged before negotiating security keys. Therefore, attackers can spoof such frames, impersonate the AP by setting up rogue devices and orchestrate several MitM attacks. For example, by spoofing the MAC address of the AP, the attacker can send deauthentication or disassociation frames to the client. Similarly, he can send a reassociation frame to the AP by spoofing the client. In either situation, the client gets disconnected from the legitimate network, resulting in DoS attacks. To counter these issues, the PMF standard was introduced.

### *2.3. Protected Management Frames (PMF)- IEEE 802.11w*

The PMF or IEEE 802.11w standard was ratified in 2009 and became a part of 802.11i standard in 2014 (Wright, Charles V., Fabian Monrose, 2009). Although PMF has been around for a longer time, its market adoption was relatively low as it was an optional feature for existing WPA2 certifications. From 2018, WFA made PMF, a mandatory security requirement for new certifications of both WPA2 and WPA3 (Burke, 2018). When PMF is enabled, it protects some specific robust management frames, such as disassociation, deauthentication, and action frames (e.g., Spectrum Management). The two main amendments of PMF are:

1) A Message Integrity Code (MIC) is generated using the shared secret IGTK (Integrity Group Temporal Key) that encrypts broadcast or multicast robust management frames (e.g., deauthentication) for providing authentication and replay protection. MIC calculation is accomplished by Broadcast/Multicast Integrity Protocol (BIP).

2) Security Association (SA) Teardown Protection is added as an association spoofing protection mechanism to prevent spoofing attacks from tearing down an existing client association. This is accomplished with a SA Query procedure that provides protection against rogue APs or clients. It is a crypto protected probe message initiated by either party to verify the authenticity of (dis)association requests.

PMF would be effective only when both AP and client support it or every device in a WLAN supports it. However, unfortunately, most currently available WPA2 devices are not capable of this feature. Although some Cisco routers support PMF, it is rarely enabled in infrastructure networks due to enormous interoperability issues. It is also almost non-existent in IoT devices due to resource-intensive crypto operations. On the other hand, even though PMF ensures data origin's authenticity of specific robust management frames such as deauthentication or dissociation frames, it does not protect other pre-authenticated management frames, such as beacons, probe responses, authentication, or (re)association frames (Bertka, 2012). This fundamental conrundum still challenges the security of not only WPA2 but also WPA3 devices, and it allows attackers to introduce MitM attacks.

In the next section, we outline the procedures involved in performing MitM attacks and analyze MitM attack´s capabilities in manipulating wireless traffic in a WLAN.

### *2.4. Fundamentals of MitM attacks in Wi-Fi networks*

According to (Conti et al., 2016), a MitM attacker in Wi-Fi networks can eavesdrop on the wireless communication between two end devices and, in some cases, can even actively manipulate the data flow. To successfully implement MitM attacks in Wi-Fi networks, attackers follow general procedures, as shown in Fig.2 (Kaplanis, 2015). During the first stage, information-gathering, the attacker may devise wardriving tools (e.g., Kismet) to deduce useful identifiers (e.g., SSID, MAC address, and channel) about the AP and clients in a WLAN. Using the deduced information from this stage, the attacker sets up a rogue AP (also known as Evil-Twin) for masquerading as the real AP in the second stage, which is instrumental in achieving the MitM position.

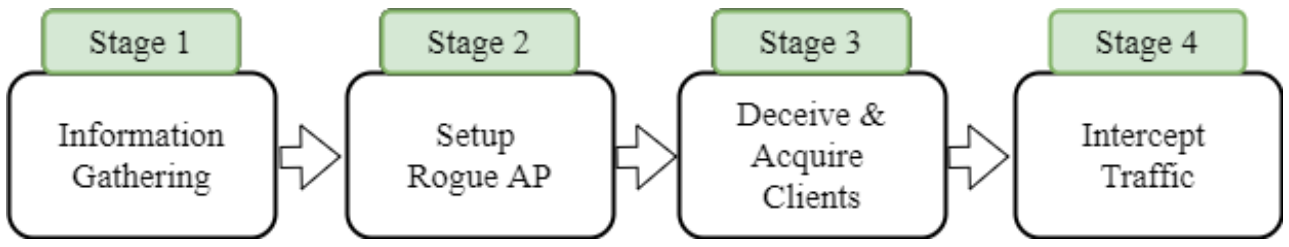

**Fig. 2. General procedure for MitM attacks.**

In the third stage, the attacker tries to deceive the clients in a WLAN. To do this, firstly, the rogue AP transmits the strongest Wi-Fi signals to lure the clients and waits for any clients who accidentally connect to the rogue AP so that he can begin capturing their traffic. He also plans for a series of active attacks (e.g., deauthentication or disassociation attack) to disrupt communication to force clients (victims) to connect to the rogue network. Once victims get connected to the rogue AP, the attacker can actively intercept traffic in the final stage. In the next section, we analyze how rogue AP-based MitM attacks manipulate protected or encrypted link-layer traffic between a client and AP in a WLAN.

#### *2.4.1. Rogue AP based MitM attack and protected Wi-Fi networks*

Usually, rogue APs are devised to trick the client into connecting to separate networks other than real AP in a WLAN (Alotaibi & Elleithy, 2016). Here, we consider a rogue AP scenario (which we will refer to this as a traditional rogue AP MitM attack from now on) where the attacker acquires MitM position between the real AP and client, as depicted in Fig.3. We also assume that the attacker knows the Wi-Fi passphrase as he is already connected to the real AP.

To acquire the MitM position, the attacker usually introduces his rogue device (e.g., laptop) between the client and real AP that generally operates with two Wi-Fi cards, a built-in card integrated into the device, and another one that can be a plug-and-play wireless card or a USB dongle. The plug-and-play card acts as the rogue AP by spoofing the real AP to the client, while the built-in card is usually associated with the real AP (Alotaibi & Elleithy, 2016; Roth et al., 2008).

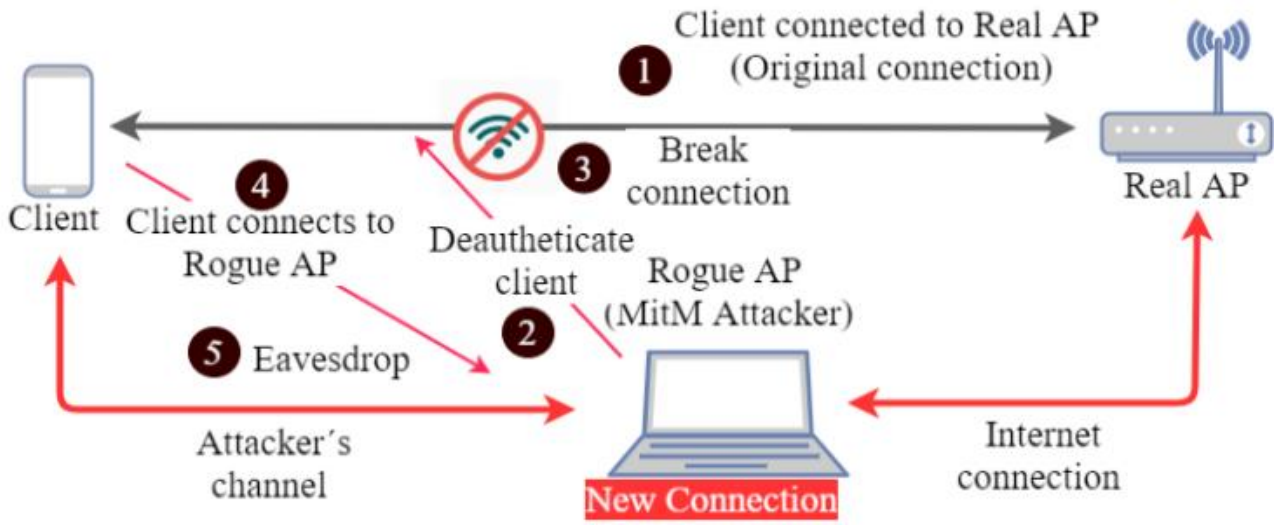

**Fig. 3. Traditional rogue AP based MitM attack.**

Then the attacker creates a rogue protected network with the same SSID, MAC address, and known security key (Wi-Fi passphrase) used in the real network to trick the user into connecting to the rogue AP's network. Wireless packets are relayed between the plug-and-play card and built-in card using a bridged network connection or traffic forwarding for providing Internet connectivity to the victim. An example of a traditional rogue AP MitM attack can be found in (Yeahhub, 2018).

We note that this traditional rogue AP MitM attack deletes the client's legitimate security association (original connection) with the real AP and forces it to perform a new authentication and association using a Wi-Fi passphrase with the attacker's rogue AP. This implies that a Wi-Fi passphrase must be known in order to perform such MitM attacks. Moreover, two separate session keys are generated since the plug-and-play (rogue AP) and the built-in card form two separate wireless connections between the client and the real AP. Therefore, the attacker cannot manipulate link-layer traffic between the client and the real AP. However, once the MitM position is acquired, the attacker usually intercepts or manipulates the Internet traffic (between the client and web server) provided by the bridged connection or traffic forwarding between the plug-and-play card and built-in card. On the other hand, the bridged connection cannot be used to block or inject protected link-layer frames between the end devices. Most importantly, traditional rogue AP MitM attacks will not be successful if PMF is enabled. This is because spoofed deauthentication will be ignored while disconnecting the existing connection.

In contrast, MC-MitM attacks, our research focus, acquires the MitM position efficiently between an already connected client and the real AP without possessing a legitimate Wi-Fi passphrase and deleting the original security association between them. Moreover, the use of different channels enables such attackers to cleverly spoof end devices and actively manipulate the encrypted link-layer traffic of a single connection between the client and the real AP. MC-MitM attacks can also acquire MitM positions in PMF environments.

## 3. Technical setup and inner workings of Multi-Channel MitM attacks

In this section, we elicit the technical setup and inner-workings towards acquiring the MC-MitM position between Wi-Fi devices. Our main aim is to evaluate the capabilities of MC-MitM attacks in manipulating protected Wi-Fi networks. We compare the characteristics of MC-MitM with traditional rogue AP-based MitM attacks in Wi-Fi networks. Finally, we analyze how MC-MitM attacks become possible in WPA3 networks and related issues.

### 3.1. Overview of MC- MitM attacks

Vanhoef et al. introduced the MC-MitM attacks against protected Wi-Fi networks in 2014 (Vanhoef & Piessens, 2014). In this kind of attacks, the main goal of the attacker is to obtain a MitM position between two already connected wireless devices without breaking their original security association and then to forward or exchange encrypted frames between them reliably. Once the attacker acquires this MitM position, he can effectively manipulate wireless frames in a way that is entirely legitimate to the victims. There are two prominent advantages in using a MC-MitM position: (1) victims remain unaware of the attack since their original connection or current security association is not disturbed; (2) attackers can bypass new authentication and association with the real AP (Chi et al., 2020). The latter one is more significant as the attacker does not hold a pre-shared Wi-Fi passphrase, which is the main parameter for deriving the session key during a 4-way handshake. To enter the network using the MitM position, the attacker uses two different channels (therefore, named as Multi-Channel-MitM) to simultaneously communicate with both the client (victim) and real AP, as shown in Fig.4.

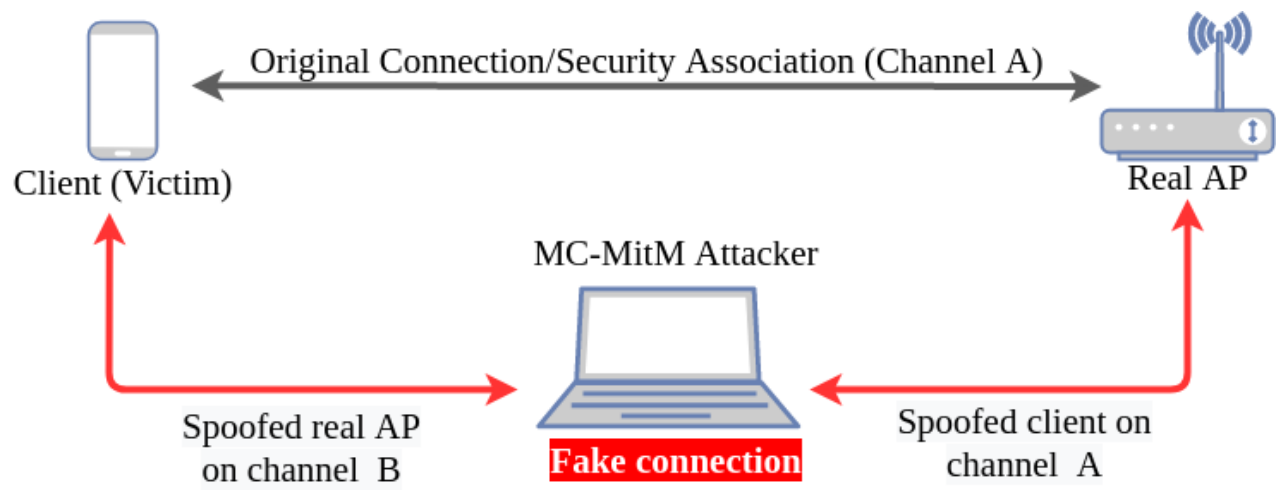

**Fig. 4. MC- MitM attack.**

As shown above, in the MC-MitM setup, the attacker cleverly spoofs communicating end devices (client and real AP) respectively on an on-side channel, which eventually drives the end devices to negotiate the same session key during a 4-way handshake mechanism. Besides, the attacker ensures that the client and real AP will never communicate directly through their original connection as he exchanges the specific communication between end devices with the help of a fake connection formed using two different channels. However, acquiring the MC-MitM position is a complicated procedure in protected wireless networks due to the 4-way handshake mechanism. The reason is that the attacker has to manage negotiated session keys derived from parameters, including the MAC address of the AP and client while maintaining the current or original security association between them. To carry through these conditions, a MC-MitM attacker performs the following two intriguing procedures: 1) Setup rogue interfaces for spoofing the victims; 2) Force the victims to switch to rogue channels. We demonstrate the inner workings of these two procedures, respectively, in Sections 3.2 and 3.3.

### 3.2. Rogue interface setup for spoofing the victims

This section demonstrates how the MC-MitM attacker sets up rogue interfaces for obtaining a MitM position between a legitimate connection, as shown in Fig.5. Then Fig.6 depicts how this legitimate connection is tweaked into the MC-MitM attack setup using spoofed interfaces. At first, the attacker inserts a laptop with a dual interface setup that simultaneously clones the targets, i.e., a real AP and a client (e.g., mobile devices, laptops, tablets) on different channels.

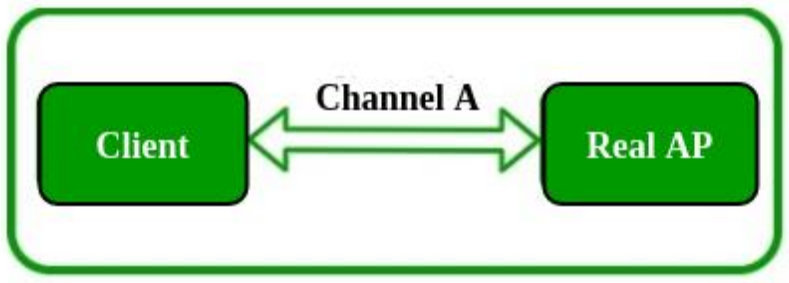

**Fig.5. Legitimate connection.**

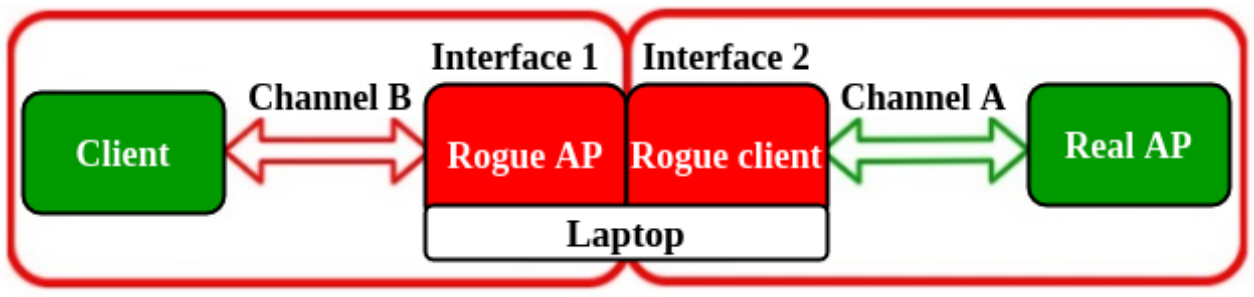

**Fig. 6. MC- MitM attack setup.**

On the one hand, the first interface clones ESSID (Wi-Fi network name), MAC address, and other necessary parameters and spoofs the real AP for the client. On the other hand, the second interface spoofs the client by cloning the client's MAC address for the real AP. These two interfaces (with Wi-Fi antennas) must be in a physically reachable range (preferably 1-2 meters) to effectively relay frames between different channels. The real AP is now cloned on a channel (channel B) other than the real channel (channel A) to connect with the client. This is an essential requirement because using the same MAC address as the real AP on the same channel (channel A) is impossible since the targeted client and real AP are already communicating with each other. Moreover, the rogue client (interface 2) needs to work on the same channel (channel A) of the real AP to show its presence on the real channel itself. Finally, to manage acknowledgment frames (ACK), the attacker modifies the firmware of interface 2 such that the rogue client will send ACKs when it receives unicast frames from the real AP. Once masquerading is complete, the rogue client (interface 2) can listen on channel A for the real AP, while the rogue AP (interface 1) listens on channel B for the client. Cloning two different interfaces in this way allows the interfaces to copy and exchange all frames from one channel to another, which drives both the client and real AP to negotiate the same session key during a 4-way handshake process. In the next section, we explain how end devices negotiate the same session key.

#### 3.2.1 Steps to negotiate the same session key

As mentioned previously, since the MC-MitM attacker does not delete the original security association between the client and real AP or does not create a new connection using a Wi-Fi passphrase, the client and real AP continue to maintain their security association (current state machine) and retain details about PMK (a hash value derived from Wi-Fi passphrase and SSID) and association identifiers (association ID) in the state tables stored in the cache of their Wi-Fi chips (Frankel et al., 2007). To acquire the MitM position, the attacker first forces the client to connect to it. While forcing the client, the attacker transmits already collected beacons of real AP. When the client sees such beacons on channel B, it recognizes that the network is already authenticated or connected (as per the preferred network list) and sends a probe request with a selected SSID. Consequently, the attacker's rogue AP sends a custom probe response to the client on channel B, making the client to send an authentication request frame to it. At this moment, the rogue AP collects that authentication request frame and retransmit it on channel A using the rogue client. The real AP accepts it, and in response, it sends an authentication response frame on channel A, which the rogue client collects and retransmits on channel B. In the same way, association frames are exchanged.

Following a successful association, the real AP initiates the 4-way handshake. At this moment, as explained above, the rogue client and rogue AP setup collects each handshake message from its originating channel and retransmit it on another channel. Even though handshake messages are exchanged between two different channels, they will have a valid MIC (from message 2) when processed by the real AP. As a result, the client and real AP derive a new and same PTK (session key) on respective sides. Moreover, the real AP honors all these exchanged frames. This is because 1) real AP remembers the client's original security association; 2) frames are transmitted on the same operating channel (channel A) of the real AP. Once the session keys are negotiated, the attacker manages all the communication (data frames) between end devices through his MC-MitM setup so that he can reliably block, delay, buffer, modify, inject, or replay encrypted wireless frames. In this way, the attacker bypasses the need for new authentication and association using the Wi-Fi passphrase and achieves the MC-MitM position.

Although the MC-MitM position forces end devices to negotiate the same session keys, the attacker cannot acquire those keys as he is merely exchanging encrypted frames between two channels. Therefore, a MC-MitM position cannot decrypt any traffic passing through it on its own. Instead, the attacker employs the MitM position to exploit specific known vulnerabilities in WPA or WPA2 to potentially decrypt wireless traffic, as concisely discussed in Section 3.4.

### *3.3. Forcing the victims to switch to rogue channels of interfaces*

As far as the real AP is concerned, it always transmits and receives frames on its operating channel (channel A). In the previous section, we explained how the MC-MitM attack setup uses interface 2 (rogue client) to communicate with the real AP and retransmit its frames on another channel (channel B) using interface 1. In this section, we demonstrate how the attacker uses rogue AP (interface 1) to force the client to connect to the rogue AP on its channel without deleting its current security association with the real AP. In terms of forcing clients to connect to the attacker's channel, we divide MC-MitM attacks into two variants: (1) base variant and (2) improved variant. We use these types later to classify existing multi-channel attacks in Section 4.

#### *3.3.1 Base variant.*

This is the first MC-MitM attack variant presented by (Vanhoef & Piessens, 2014). In this variant, the attacker first constantly jams the original channel (channel A) of the real AP until the targeted client connects to his rogue channel (see Fig.7). This is accomplished with commodity hardware capable of jamming Wi-Fi frames on a specific channel. Due to jamming, the client loses connection from real AP that is on the real channel (channel A). Meanwhile, the attacker with the rogue AP advertises beacons on rogue channel (channel B) to trick the victims into connecting to it. More specifically, the MitM attacker copies beacons of real AP from the real channel and retransmits them on the rogue channel.

Note that the jamming does not break the original security association. Instead, it just makes target networks unavailable for some time. As per the 802.11 standards, a client will always choose an available network or a network with the strongest signal. Therefore, victim switches to the rogue AP's channel and starts transmitting data on it. Additionally, the attacker observes specific probe requests from the client and instantly replies with custom probe responses to force it to switch to his channel. As soon as the client switches to the rogue channel, the attacker stops jamming.

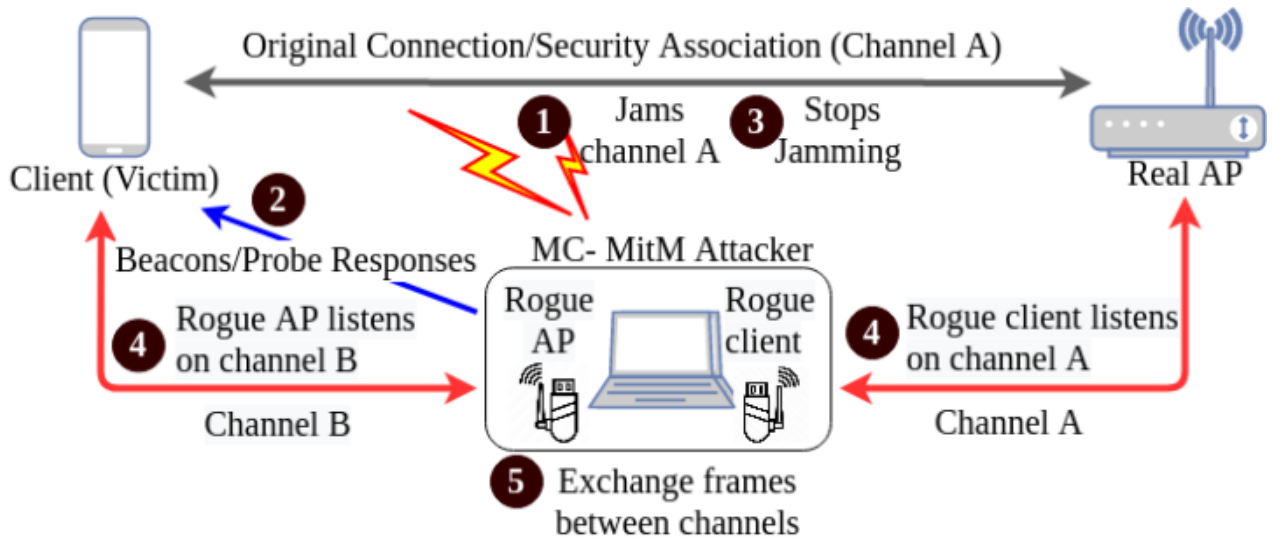

**Fig. 7**. **Multi-Channel MitM Attack- Base variant.**

As of now, the attacker acquires the MitM position, and he starts exchanging frames between the client and the real AP. This base variant can also attack PMF capable devices because management frames such as beacons or probe responses are not protected even if PMF is enabled (recall Section 2.3). We implemented and tested the base variant by using the Modwifii tool (Vanhoef, 2015).

#### *3.3.2 Improved variant.*

This variant appeared with several improvements over the base variant and was also proposed by (Vanhoef & Piessens, 2018). With this improved variant, the MC-MitM attacker uses channel switch announcements (CSA) to trick the client into connecting to his rogue channel. CSAs can be transmitted by inserting a CSA information element inside beacon frames, probe response frames, and action frames. Like the base variant, the improved variant also sends beacons or probe responses with spoofed CSA to trick both regular and PMF clients, whereas action frames are protected when PMF is enabled. Once the client receives CSAs, it will instantly switch to the rogue channel so that the attacker starts exchanging frames between the client and the real AP. Fig.8 demonstrates the working of MC-MitM improved variant attacks.

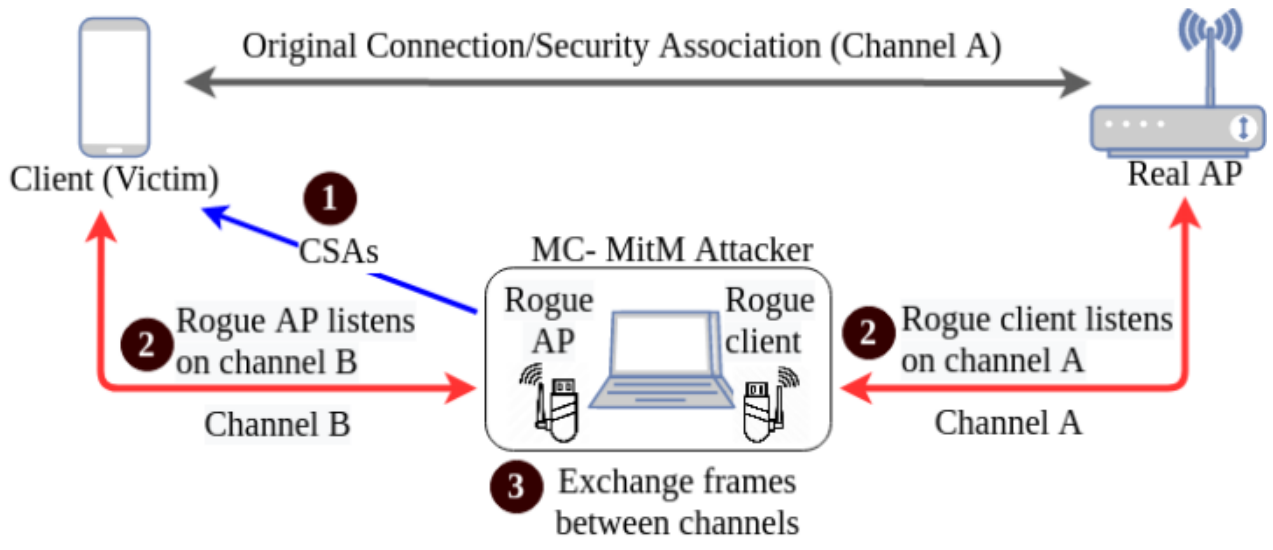

**Fig. 8**. **Multi-Channel MitM Attack- Improved variant.**

Utilizing CSAs is reliable and does not disturb the current security association. It is a normal activity of APs in certain conditions (e.g., noise or congestion) that the clients cannot decline. Moreover, only a few CSAs (max 4 or 5) are enough to force the victims for the desired channel change. Fig.9 depicts the structure of the CSA information element. The channel switch mode field regulates whether a wireless client can continue (when the mode is 1) or stop (when the mode is 0) sending information on a particular channel. The new channel number field indicates the expected channel to which the clients must go. The channel switch count field represents the remaining number of beacon interval to wait (a zero value indicates that channel switch is imminent) for a client before a channel switch.

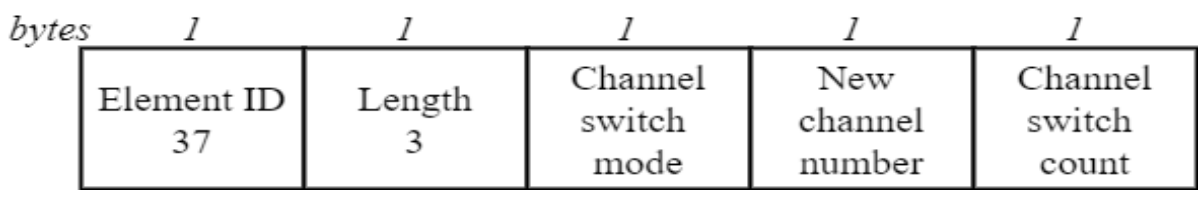

**Fig. 9. Structure of a CSA element** (IEEE 802.11 Standard, 2012)**.**

**S**ince CSAs can instantly switch channels of clients, channel jamming is not required in this variant, which considerably decreases attacker's efforts. Additionally, the attacker can spoof CSAs to the client to switch back to the real channel after abusing it. In any case, the client (victim) remains unaware of the attack. All combined, the MC-MitM improved variant increases impacts of attacks. We implemented and tested the MC-MitM improved variant attack by using the MitM channel package (Lucas Woody, 2018). In Table 1, we summarize the features of MC-MitM attack variants.

**Table 1 - Comparison of MC-MitM attack variants**

| **Characteristics** | **Base variant** | **Improved variant** |
|---|---|---|
| Employ beacons | Yes | Yes |
| Employ probe responses | Yes | Yes |
| Employ action frames | No | Yes |
| Needs jamming to launch attack | Yes | No |
| Ability to attack PMF clients | Yes | Yes |
| Cost effective method | No | Yes |
| More reliable method | No | Yes |
| More impactful method | No | Yes |

### *3.4. Decrypting Wi-Fi Frames using the MC-MitM position*

According to the IEEE 802.11 standard (Hiertz et al., 2010), performing decryption and encryption of Wi-Fi frames requires generating the session key (PTK) during a 4-way handshake mechanism. In section 3.2, we have seen how MC-MitM attackers manage to force end devices to negotiate the same session key without possessing a pre-shared Wi-Fi passphrase. Furthermore, we indicated that a multi-channel attacker has no access to those negotiated keys. This is a significant challenge because decrypting frames requires the knowledge of a particular session key. Even so, the MC-MitM attacker can achieve the above challenge in many ways. In previous MC-MitM attacks on WPA, the attacker abused specific weaknesses in encryption algorithms (e.g., MIC key derivation vulnerability in TKIP) so that he would be able to decrypt wireless frames (Vanhoef & Piessens, 2014). However, such decryption technique was a hard-to-win race condition since the attacker had to predict several parameters; moreover, he could decrypt only some arbitrary frames. On the other hand, with the disclosure of key reinstallation vulnerabilities in WPA2 standards, MC-MitM attackers could decrypt comparatively large numbers of packets in a short period irrespective of data confidentiality protocols used in Wi-Fi networks. Therefore, we show how key reinstallation vulnerabilities presented in (Vanhoef, 2017b) allow attackers to decrypt Wi-Fi frames of a particular communication session between end devices.

Regarding the key reinstallation vulnerability, the major flaw is in the WPA2 standard that makes every Wi-Fi capable device reset nonce and packet counters of data confidentiality protocol. This happens automatically whenever a session key (re)installed on the client-side during a 4-way handshake. This means that clients are already reusing nonce values even without an attacker being present. Figure 10 depicts how encryption works generally in a Wi-Fi network.

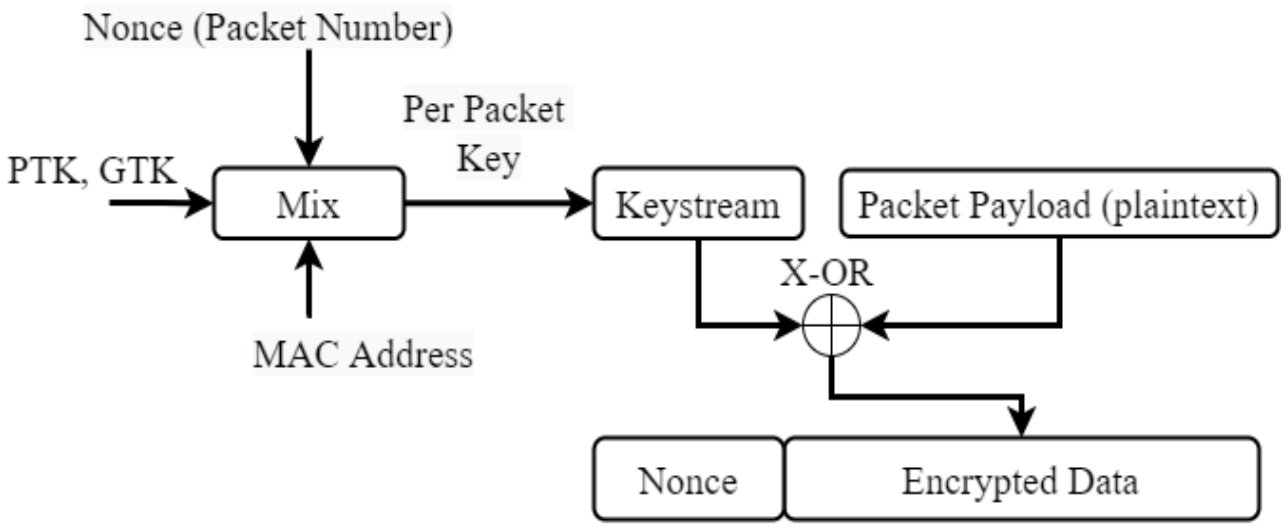

**Fig. 10. Generalized encryption procedure in Wi-Fi.**

As per Fig.10, once the session key is negotiated, it will be combined with the transmitter´s MAC address and the nonce value (packet number), which is incremented by one for every transmitted frame, and eventually a unique per packet key is derived (Vanhoef, 2018). This per packet key is fed into a stream or block cipher (encryption algorithm) to generate the corresponding keystreams and is then XORed with the plaintext packet payload to create the ciphertext or encrypted data corresponding to a particular frame. Finally, the nonce value is also appended to the header of the frames so that the receiver will be able to decrypt the frames. In this way, a nonce value is used to form a unique per packet key. An essential requirement here is that under a particular session key, the nonce value should only be used once. If the encryption algorithm ever reuses a nonce value, it will generate the same per packet key and yield the same keystreams. This is the major vulnerability that is wisely exploited by MC-MitM attackers to decrypt the Wi-Fi frames effortlessly.

Fig.11 shows the technical representation of how a MC-MitM attacker can decrypt Wi-Fi frames. During stage 1, the MitM attacker exchanges the first three handshake messages between channels without any modifications. Actual MitM attack will start from stage 2, where the MitM attacker blocks the message 4 from the client and does not forward it to the AP. From the client's perspective, the handshake is completed, and so it installs the session keys (PTK and GTK) and initializes its nonce and replay counter values to zero as per Wi-Fi standard. Since the AP has not received message 4, it retransmits message 3 to the client in order to continue the handshake progress, which will be then forwarded by the MitM attacker to the client. Note that, as per 802.11 standards, if the AP does not receive message 4 because of reasons like noise or congestion in the network, it will always retransmit message 3. Consequently, the client sends message 4 (with a nonce value one as incremented due to the new frame) and is in encrypted form since the client has already installed the session key. Following the sending of message 4, the client again installs (reinstalls) session keys.

When a key is reinstalled, the nonce (packet number) and replay counter values are reset to zero. This means that if the client sends another data frame, it will again use the old nonce value one and thus uses an already-in-use session key (per packet key) for encrypting the data frames. Reusing the nonce in messages (as shown using green arrows in stage 2) causes the same keystreams to be reused.

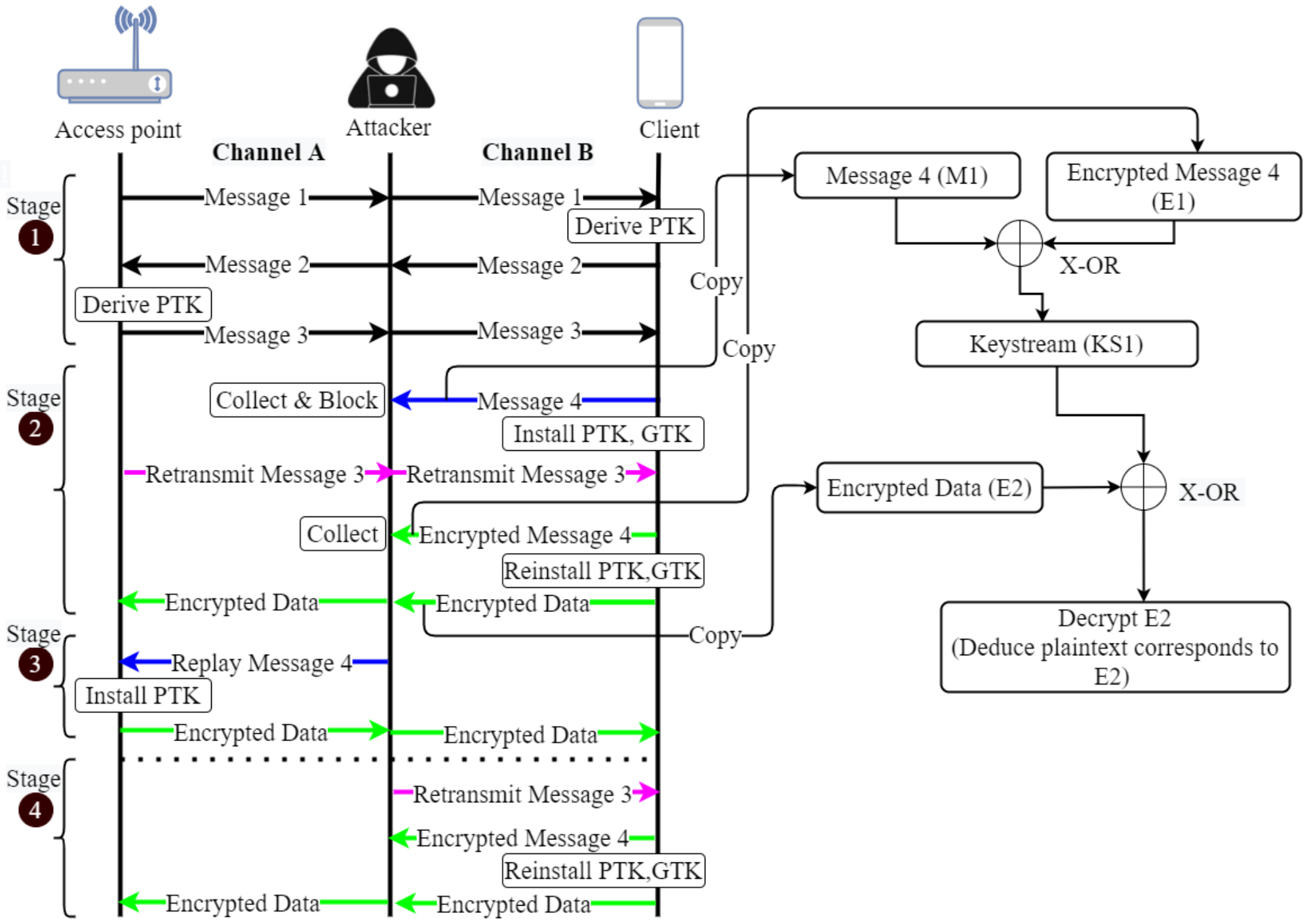

**Fig. 11. Key reinstallation attack using the MC-MitM position.**

At this stage, the multi-channel attacker starts abusing the nonce reuse scenario to recover the keystreams corresponding to the nonce value one. To do so, he performs the following: first, the attacker copies message 4 (M1) and encrypted message 4 (E1) to X-OR them to learn the keystream (KS1) belongs to the nonce value one. The reason for this is, the X-OR operation between plaintext and its encrypted message gives the keystream for that encryption. Second, the attacker copies encrypted data frame (E2) in stage 2, which also uses nonce value one due to key reinstallation, meaning that it might have used precisely the same keystream (KS1). Finally, the attacker again performs an X-OR operation between KS1 and E2 to get the plaintext corresponding to the encrypted data (E2).

In stage 3, the MitM attacker replays the already collected message 4 (blue arrows) towards the AP. The AP then accepts this replayed message 4 since the replay counter was reset during the last key reinstallation and begins sending encrypted data using an already-in-use session key, which can also be decrypted easily by the MitM attacker. Similarly, in stage 4, the attacker can force key reinstallations by replaying message 3 (pink arrows) retransmitted during stage 2. We remark that a key reinstallation attack enables decryption of just an encrypted message with a nonce value one. To decrypt the client's subsequent messages, the attacker must replay message 3 to induce nonce and replay counter reset during the key reinstallation attacks. By forcing the key reinstallations continuously in this manner, the attacker can decrypt a greater number of Wi-Fi frames. These frames can be a part of some TCP connections when user crawls websites or exchanges personal data.

Interestingly, (Chi et al., 2020) showed how MC-MitM attackers could capture Wi-Fi frames between two legitimate devices and then directly decrypt Wi-Fi frames on-the-fly using an open-source library, such as pyDot11. In this case, the attacker holds the pre-shared Wi-Fi passphrase, meaning that he is an insider and tries to decrypt a particular communication session between clients and the AP in the same WLAN.

### *3.5. Obtaining Wi-Fi data using the MC-MitM position*

With the advent of recent aggregation and fragmentation security vulnerabilities found in the 802.11 standards (Vanhoef, 2021a), MC-MitM attacks become more widespread and practical to trigger FragAttacks towards WPA, WPA2, and the new WPA3 networks. FragAttacks enable the attackers to legitimately inject specific Wi-Fi frames and then obtain user's sensitive data. In this subsection, we show how FragAttacks leverage the MC-MitM position to intercept and obtain user´s sensitive data.

In Wi-Fi, sending small Wi-Fi packets is inefficient because every frame must have a header and separately acknowledged, which may often induce high overhead on Wi-Fi chips. Therefore, small packets are aggregated into a larger frame containing multiple packets with the frame aggregation feature. Every Wi-Fi frame header contains an aggregation flag that indicates whether the frame payload contains a single (normal) packet or multiple aggregated network packets. Nevertheless, the major flaw is that the frame aggregation flag in the Wi-Fi header is not authenticated. This allows the attacker to flip the respective flag and trick the victim into processing encrypted frames by injecting frames towards him. Fig.12 illustrates the aggregation attack in Wi-Fi.

During stage 1, the attacker acquires the MC-MitM position between the client (victim) and AP. He also sets up a fake DNS and web server for impersonating websites and Internet access for the client. In stage 2, the attacker tricks the client into connecting to his web server. This is accomplished by sending an email to the client, and when clicked, it causes downloading an image from the attacker's web server, establishing a TCP connection with the web server. The attacker manages this TCP connection to send a malicious TCP packet (IPv4 packet) to the client (stage 3). In stage 4, the AP encrypts the injected IPv4 packet as a normal frame and forwards it to the client. Afterward, the MC-MitM attacker subsequently identifies this frame and flips the aggregated flag before forwarding it to the client (stage 5).

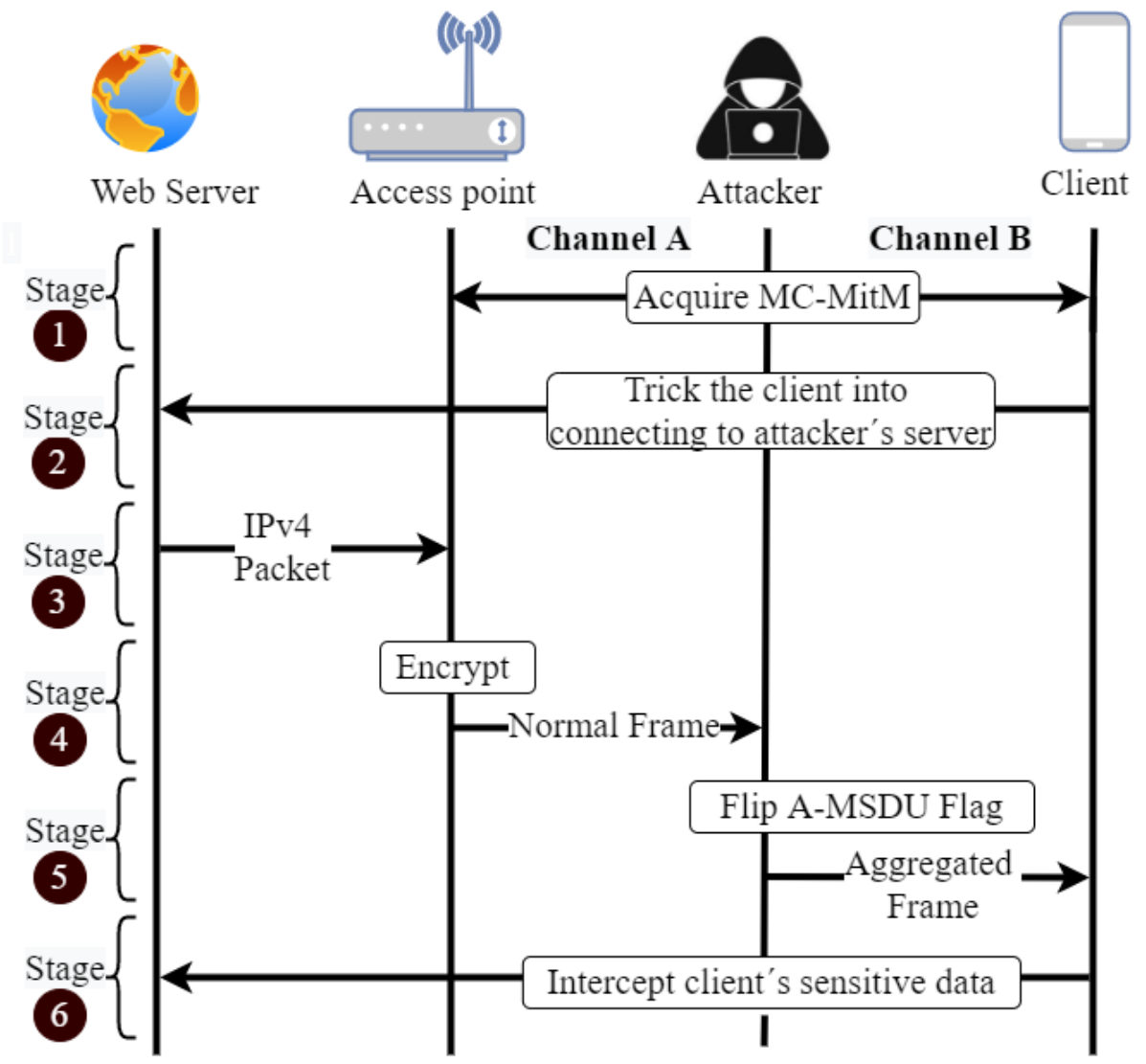

**Fig.12. Aggregation attack using the MC-MitM position.**

On the other hand, the client will not detect this aggregated flag due to the design flaw. Therefore, the frame becomes an A-MSDU (Aggregate MAC Service Data Unit) frame so that the attacker can inject IP packets as subframes. When the client processes such aggregated frames, it will be tricked into connecting to the fake DNS server. The injected IP packets can be ICMP router advertisements or DHCP packets. Once the client is connected to the attacker's DNS server (stage 6), he can intercept all the client's IP traffic and obtain sensitive data (e.g., log-in details), especially while using insecure websites.

Regarding Wi-Fi fragmentation, it is the process by which large frames are split into smaller frames in order to avoid the chances of frames being corrupted. This process also facilitates the retransmission of specific lost frames whenever required. There are two significant flaws discovered concerning the fragmentation allowing the revelation of victim data.

1. Even though the fragments of a frame are encrypted using the same key, there is no verification procedure (ensuring whether the same key encrypts the received frames) at the receiver. The sequence number field in a fragment is also not authenticated. As a result, attackers can abuse the lack of verification to inject and mix frames with different keys (with previous sequence numbers), which will be reassembled by the receiver (see Fig. 13).
2. The fragment cache is not cleared when clients (re)connect to particular Wi-Fi network. Therefore, this flaw allows the attacker to inject frames into the fragment cache, which will be reassembled with the client´s fragments (see Fig. 14).

As shown in Fig.13, the attack starts with acquiring the MC-MitM position during stage 1. In this stage, the client is first tricked into visiting an attacker-controlled website. For example, the attacker may send phishing emails, show third-party advertisements, or posts on blogs the client may visit, by social engineering the user activities, and load the corresponding Internet resource (web pages) on the attacker´s web server. The main goal of this step is to create an attacker-destined packet (i.e., a packet with the destination IP address, which in this example is 3.5.1.1). When the client visits such long web pages or URL, the resulting packets will be split into two fragments (Frag 0 and Frag 1) as highlighted using green arrows. Note that fragments of the same frame will have the same sequence number *s* and incremental packet number *n*, and the session key *k* encrypts the fragments. The sample contents of these fragments are shown using dotted red arrows. The MC-MitM attacker detects and collects these fragments according to their unique length and only forwards the first fragment (Frag 0) to the AP. Upon receiving this fragment, the AP decrypts this fragment and stores it in its cache or memory.

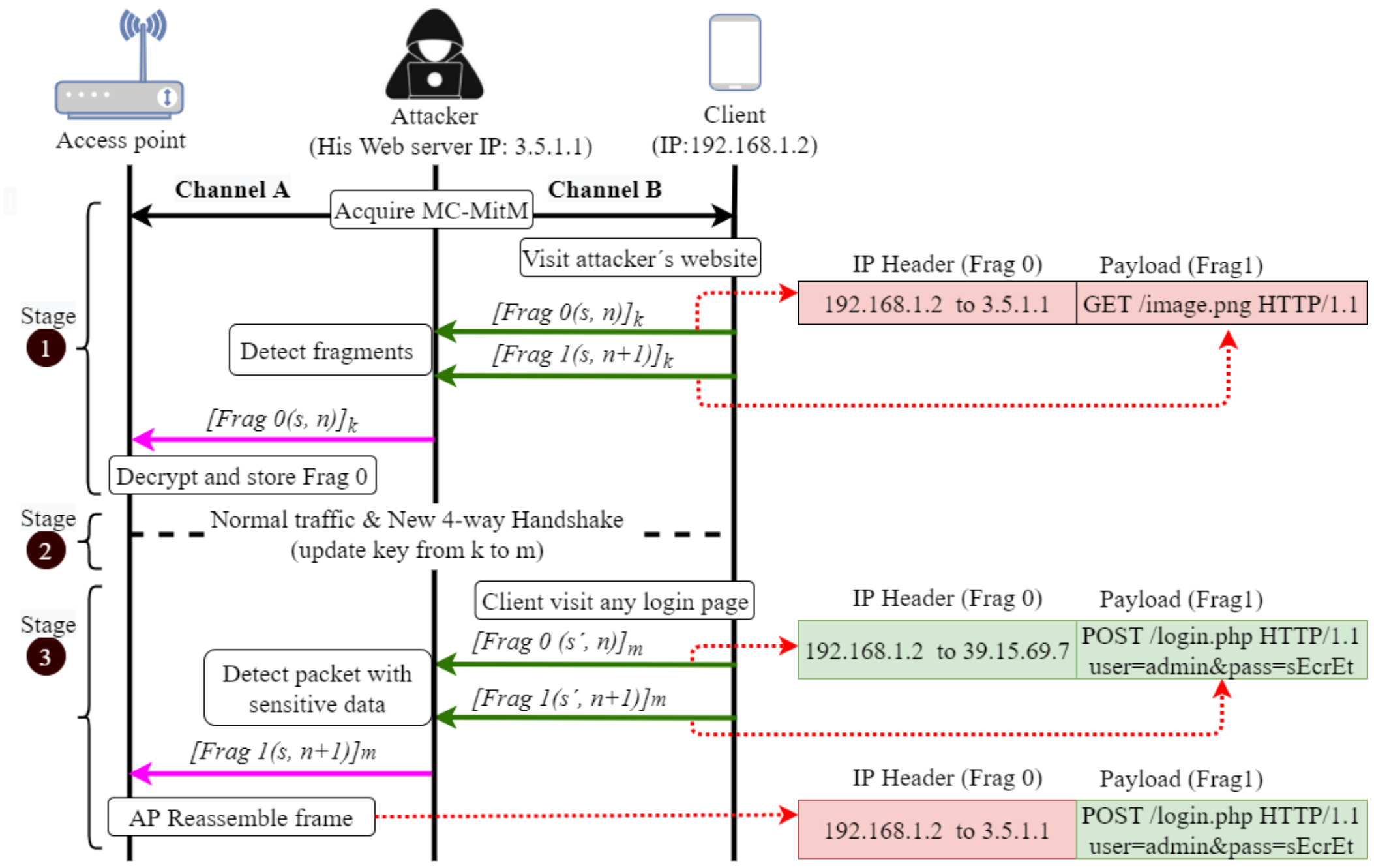

**Fig. 13. Fragmentation mixed key attack using the MC-MitM position** (Vanhoef, 2021a).

During stage 2, the attacker forwards all other normal traffic without the packet number to ensure that the first fragment is never removed from the AP´s cache. He also waits for session key renewal after a 4-way handshake. The attacker can predict the rekey as it occurs in regular intervals. By rekeying, packet numbers of the encryption protocol will be reset to zero. With these assumptions, stage 3 begins. Here, when the client visits any web page (having IP address, which in this example is 39.15.69.7) and provide log-in details through that particular web page, the corresponding IP packet (i.e., HTTP POST request) is split into two fragments and encrypted using the new session key *m*. At this moment, the MC-MitM attacker identifies those fragments with sensitive data and only forwards the second fragment (Frag 1) by tweaking its sequence number to *s* as the one used with the first fragment sent during stage 1. However, the AP combines both fragments (highlighted using pink arrows) into a new reassembled frame due to the design flaw. Since the attacker-destined packet is now combined with the user´s sensitive data, it will be sent to the attacker´s web server.

In the fragmentation cache attack, as illustrated in Fig.14, the attacker basically targets enterprise networks such as eduroam, where each user has a unique username and password. In the first stage of the attack, the attacker accesses the network with his credentials. He also waits for an authentication request from the client (victim) and immediately injects an IP packet (Frag 1) to the AP by spoofing the client's MAC address. The goal of this packet injection is to create an attacker-destined packet with a destination IP as 3.5.1.1. Consequently, the AP decrypts this attacker-destined packet and keeps it in its memory or cache with the client's MAC address. Then the attacker leaves from the AP with a deauth frame. Due to the design flaw, the attacker-destined packet remains in AP's cache even though the attacker disconnects from the network.

In stage 2, the attacker oversees that he never sends any frame to AP with sequence numbers to ensure that the injected frame is not cleared from its cache. He also waits for the client to normally connect to the AP by using valid credentials. In stage 3, soon after the client connects to the AP, the attacker establishes the MC-MitM position between them. The attacker now waits for the client to visit any web page and formation of two fragments (as illustrated in stage 3 of Fig.13). Ultimate goal of the MC-MitM attacker here is to identify fragments (Frag 1 and Frag 2) and only forwards the second fragment to the AP. Afterward, the AP wrongly reassembles the previously injected attacker-destined packet with number *n* during the stage 1, with the newly forwarded packet with number *n+1* (both highlighted as pink arrows) into a single frame. This happens because the identities like MAC address and sequence number of fragments are the same. Therefore, the design flaw exploited here is that the AP does not maintain identities of fragments in enterprise networks when users (re)connect to a particular network. Eventually, the reassembled frame is sent to the attacker's server, revealing the user-sensitive data. The fragmentation cache attack is also possible against clients. In such cases, the attacker injects malicious IP packets towards clients to trick them into connecting to a fake DNS server even if clients are associated with trusted personal networks, such as home networks, coffee shops, where the Wi-Fi passphrase is publicly known to everyone.

We emphasize that the revelation of data is possible with FragAttacks as long as the client uses insecure websites that follow plaintext HTTP. Even if HTTPS is used, MitM attackers may use tools like sslstrip to bypass this upper-layer security. Therefore, nowadays, it is necessary to ensure that the website is HSTS (HTTP Strict Transport Security) compliant. Unfortunately, only very few websites dictate the use of HSTS for web communications as per the latest statistics (W3Techs, 2021) . However, the aggregation and fragmentation vulnerabilities enable the attacker to inject any packets (unencrypted packets) based on his choice into a protected Wi-Fi network and pollute communication. On the other hand, such attacks are not possible with any other tools under normal conditions.

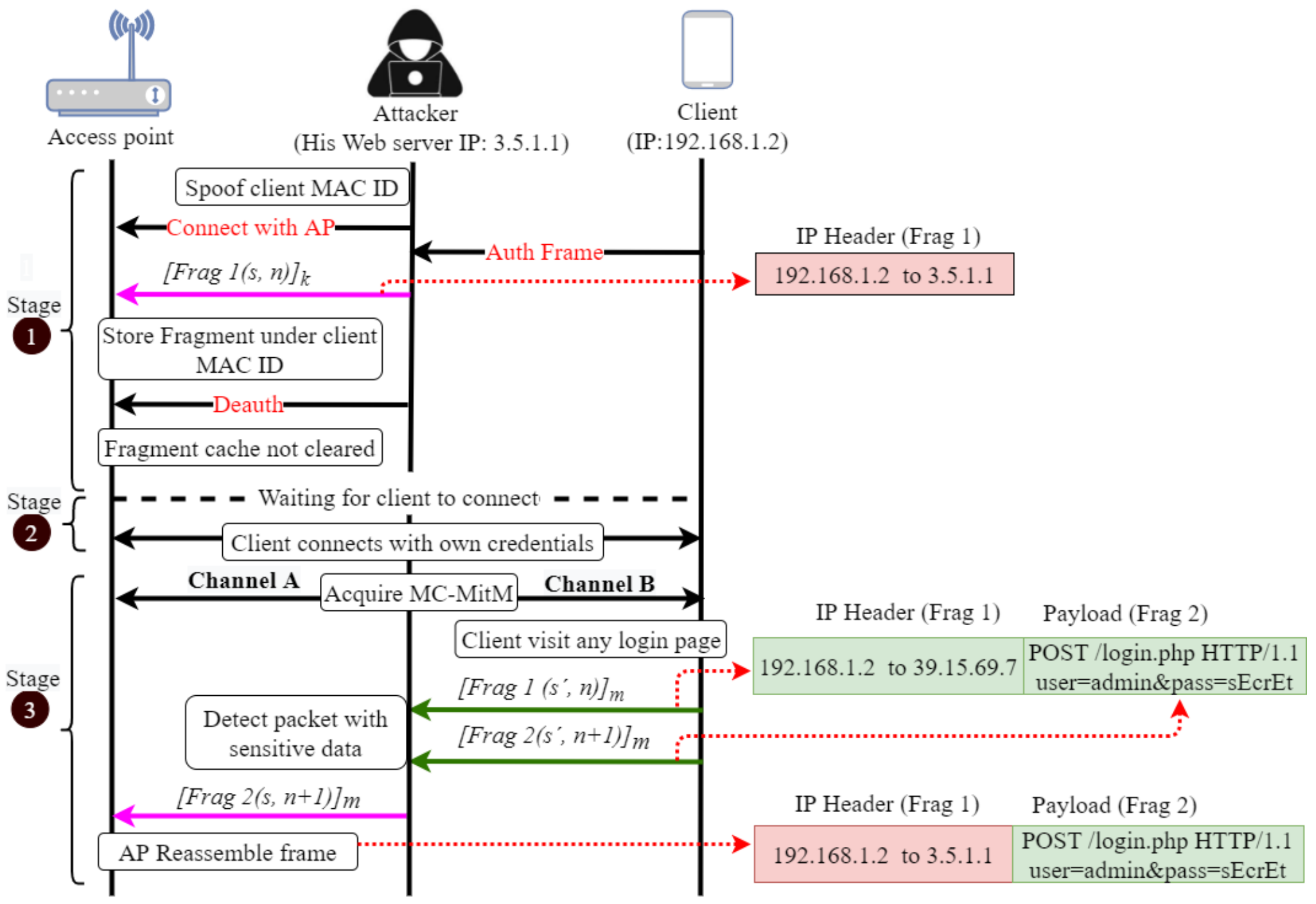

**Fig. 14. Fragmentation cache attack using the MC-MitM position** (Vanhoef, 2021a, 2021c).

### 3.6. Other special features of MC-MitM attacks

In this subsection, we present certain special features of MC-MitM attacks in manipulating protected Wi-Fi networks.

#### 3.6.1. Virtual interface support

Using a virtual interface support (a hardware technology), the rogue AP or interface 1 (see Fig. 6) employed in MC-MitM attacks can listen to multiple MAC addresses or clients simultaneously. Therefore, MC-MitM attackers can target or abuse more than one client at a time and can engender more security impact in practice (Vanhoef & Piessens, 2014).

#### 3.6.2. Detect/exploit logical vulnerabilities

The MC-MitM position can be used to detect or test any logical vulnerabilities or cryptographic implementation bugs (e.g., reusing nonces, skipping handshake messages) present in Wi-Fi handshake mechanisms (Vanhoef et al., 2017). The attackers can then exploit such vulnerabilities to perform attacks like authentication bypass, DoS, chop-chop attacks, or downgrade attacks. Key reinstallation attacks and FragAttacks are well-known attacks that exploit cryptographic vulnerabilities in various handshake mechanisms or Wi-Fi aggregation and fragmentation capabilities. On the other hand, the MC-MitM position can be used to perform traffic analysis in protected Wi-Fi networks as part of defensive security analysis.

#### 3.6.3. Jam Wi-Fi using USB dongle

The attacker uses a portable and cheap USB jammer to selectively (target specific frames) jam MAC-layer traffic on specific channels (Vanhoef & Piessens, 2014), which is comparatively difficult to be identified by using IDS systems (Gong et al., 2020). This jammer can be implemented even on a smartphone. The MC-MitM attackers appropriately use reactive or constant jamming to block or delay wireless traffic reliably.

#### 3.6.4. Legitimate behavior of the MC-MitM attacker

In both MC-MitM variants, the attacker acquires MitM position without deauthenticating the victim from the real AP. According to our analysis, the attacker does not conduct any forms of flooding attacks using spoofed beacons or any other frames while acquiring the MitM, instead, it collects the beacons of real AP and retransmits them on rogue AP´s channel. After acquiring MitM position, the attacker exchanges encrypted or manipulated frames, facilitating end devices to communicate through the attacker´s MC-MitM setup as if they are communicating with each other directly. Finally, the victim can even rejoin the real AP after withdrawing the MC-MitM position since the attacker did not force the end devices to destroy their security association.

#### 3.6.5. Trigger attacks from farther

To trigger MC-MitM attacks, the attacker need not be always close to the victim. He can use special directional antennas from farther (1 or 2 miles) and act as a repeater to obtain the MitM position and then relay the wireless frames from the AP to the victim (Vanhoef, 2018; Vanhoef et al., 2018). The attacker can also trigger attacks by cloning a far-away network and forward frames over the Internet by using specific TCP connections (Vanhoef & Piessens, 2014). However, these attacks are possible only if the attacker has prior knowledge about the network that the victim is supposed to connect. Recently, (Louca et al., 2021) demonstrate the feasibility of using channel switch announcements to acquire MitM from relatively longer distances even with the low signal strength.

In Table 2, we compare the essential characteristics of MC-MitM attacks with that of traditional rogue AP MitM attacks.

**Table 2 - Comparison of MC-MitM attacks with traditional rogue AP MitM attacks**

| Characteristics | MC-MitM | Traditional rogue AP MitM |
|---|---|---|
| Main Objective | Acquire MitM position between an already client and the AP. | Disconnect the client from the AP and create a new rogue network having the same Wi-Fi password as a real AP. |
| Num. of Interfaces | Two: for spoofing AP and the client. | Two: Spoofing as AP and connecting Internet. |
| Ability to relay | Yes | Yes |
| Ability to manipulate link-layer encrypted traffic between the client and real AP | Yes | No |
| Ability to attack PMF clients | Yes | No |
| Ability to attack multiple clients | Yes | No |
| Ability to trace logical vulnerabilities | Yes | No |
| Ability to jam | Yes | Yes |
| Behavior of the attacker | Acts as legitimate as an AP in a WLAN | Mostly acts maliciously |
| Location of the Attacker | Near to victim or far away (2 miles) | Near the victim |

### 3.7. Analysis of MC-MitM attacks in WPA3

As we highlighted in Section 3.3, since MC-MitM attack variants can circumvent PMF protection and manipulate the protected communication on WPA2 devices, MC-MitM attacks can also affect WPA3 devices. This is possible because the connection establishment process is the same in WPA2 and WPA3 except for the new Dragonfly authentication. This new authentication is merely increasing the Wi-Fi passphrase entropy and would not be a concern for the MC-MitM attacker as he does not even require it. The PMF procedures are also the same in both WPA2 and WPA3 security protocols. Therefore, the MC-MitM attacker can follow the same practices described in Section 3.2 to acquire the MC-MitM position between an already connected WPA3 client and AP. Recently orchestrated FragAttacks are the fresh examples of MC-MitM attacks in WPA3 networks.

Recently, the WFA has incorporated certain defenses against MC-MitM in their WPA3-2020 updates (Stephen Orr, 2020). New defense mechanisms incorporated in WPA3 hamper spoofing attacks, including MC-MitM attacks materializing from outsiders to a great extent. That is, as long as the attacker does not have the Wi-Fi passphrase, he cannot perform MC-MitM attacks. However, these new defense mechanisms are optional features in the WPA3, meaning that an unpatched WPA3 device (WPA3 devices that only implement mandatory security requirements such as the new Dragonfly handshake) is always exposed to MC-MitM attackers. To what extent a significant problem still needs to be explored is how to defend against various insider MC-MitM attacks effectively. This can be a significant issue when the attacker has a Wi-Fi passphrase and can access a network that hosts multiple WPA2 and WPA3 devices. In Table 3, we highlight the current issues in Wi-Fi security protocols in view of MC-MitM attacks.

**Table 3 - Current issues in Wi-Fi security protocols related to MC-MitM attacks**

| Protocol | Deauth /Dissassoc-Attacks | Outsider MC-MitM attacks | Insider MC-MitM attacks |
|---|---|---|---|
| WPA | Possible | Possible | Possible |
| WPA2 | Possible | Possible | Possible |
| WPA2-PMF | Not Possible | Possible | Possible |
| WPA3 | Not Possible | Possible with unpatched WPA3 devices | Possible even with patched WPA3 devices |

## 4. Recent MC-MitM enabled attacks in IEEE 802.11 networks and their impacts

In this section, we thoroughly review existing MC-MitM enabled attacks (attacks performed after acquiring the MitM position) towards WPA and WPA2 networks and examine whether any of these attacks can be possible in WPA3 networks. Our main aim is to study various vulnerabilities exploited and related impacts of MC-MitM attacks. To review the existing attacks, we follow the classification of MC-MitM attacks presented in section 3.3.

### *4.1. Multi-Channel MitM attacks powered by base variant*

In their work (Vanhoef & Piessens, 2014), the MC-MitM position is devised for the first time to attack the WPA-TKIP encryption protocol. They demonstrate how to abuse TKIP when used as a group cipher, targeting multicast and broadcast frames towards clients in a WLAN. While attacking a specific client, the authors employ the MitM position to block all Message Integrity Code (MIC) failure reports from other clients connected with the AP. Blocking the MIC failure report is essential to suppress TKIP countermeasures (renewing group keys for reconnection) from the AP. Further, they demonstrate how to capture and decrypt client traffic using the already known Beck and Tews method while guessing some specific frames and eventually derive the corresponding MIC key of broadcast frames. Following the MIC key's derivation, they extend the attack targeting multiple wireless clients in a WLAN. Since this MitM attack mainly exploits the flaws associated with TKIP's Michael algorithm (Beck & Tews, 2009), it can be practical in every WPA-TKIP or WPA2-TKIP network. However, this attack is not possible against WPA3 networks as the WPA3 does not support TKIP (Cisco, 2021).

A size-exposing attack has been proposed by (Goethem et al., 2016) for manipulating encrypted web traffic with the MC-MitM position while tricking the victim into sending requests and the forward frames to the real AP. This attack enables the authors to learn about the size of the resources (e.g., the size of the web packets) and then identify user web activities or websites visited. More precisely, they capture and manipulate encrypted (TKIP/CCMP) MAC layer frames of a specific TCP connection (target connection of the victim) to derive the exact size of the HTTP/S messages. The MC-MitM position helps the attacker to block unwanted background traffic (other than targeted TCP connection) to a victim and precisely calculate the size of the resources or packets accessed by him. Moreover, the MitM position manages retransmitted frames and reduces potential packet loss at the MAC layer. According to the authors, the size-exposing attack happens because the padding is never added while encrypting MAC layer frames, and no matter which encryption algorithm is used, the attacker can determine the length of encrypted plaintext in any Wi-Fi network. Therefore, such attacks are also possible in WPA3 networks.

(Vanhoef & Piessens, 2016) have presented some design flaws in random number generators (RNG) in several implementations. They illustrate how these flaws lead to predicting a group key so that an attacker can inject malicious wireless frames and potentially decrypt specific group traffic in WPA2 networks. To accomplish this, with the MC-MitM position, the attacker triggers security downgrade attacks by modifying beacons and probe responses to trick the victim into thinking that AP supports only TKIP instead of CCMP. This enables AP to start using RC4 (encryption algorithm of WPA-TKIP) for encrypting that communication session. The attacker exchanges the first two handshake messages between the AP and client with the MitM. When the AP accepts downgrade requests, it starts encrypting message 3 (containing the group key) with the RC4. The attacker then captures message 3 and recovers the group key exploiting the above-mentioned design flaws in RNG. Once the group key is derived, it enables the attacker to inject broadcast wireless packets and, in turn, decrypt all the Wi-Fi traffic.

Another security downgrade attack is presented by (Vanhoef et al., 2017). Here, with the MC-MitM position, the authors show how the attacker manipulates beacons and probe responses to trick the victim into thinking that AP supports only TKIP instead of CCMP even though both devices support CCMP. More concretely, the MitM attacker first relays messages 1 and 2 of a 4-way handshake during the attack and then blocks message 3 to hide RSNE details. Following this, the attacker sends a crafted message 1 to force the client to retransmit message 2, which will be forwarded to the AP. However, the AP wrongly interprets this message 2 as message 4 (vulnerability) and finishes the 4-way handshake. As a result, the client will connect to the AP and use TKIP as the selected cipher suite. Once accomplished, the authenticator starts encrypting frames using TKIP. As of now, the attacker can decrypt sensitive information by exploiting known vulnerabilities of RC4.

As mentioned before, since WPA3 does not support TKIP, the security downgrade attacks presented in (Vanhoef & Piessens, 2016) and (Vanhoef et al., 2017) cannot be possible in WPA3 networks.

In the mid of 2017, (Vanhoef & Piessens, 2017) have discovered severe key reinstallation vulnerabilities (nonces and replay counter reset during a session key installation) in 802.11 standards. Recall Section 3.4, where we have demonstrated the working principles of key reinstallation attacks. In practice, these vulnerabilities can be abused to decrypt TCP packets of a specific connection and then possibly hijack application layer (HTTP/S) traffic. It is also trivial for the attacker to hijack device control commands in IoT networks by replaying specific broadcast and multicast UDP packets. The KRACK was severe against TKIP and GCMP data confidentiality protocols as the attacker can even forge and inject malicious packets into Wi-Fi networks.

Like 4-way handshake abuse, MC-MitM attackers can also abuse Group Key handshake and Peer Key handshakes. Furthermore, in (Vanhoef & Piessens, 2018), which is the follow-up work of (Vanhoef & Piessens, 2017), they presented how KRACK can be performed on Tunneled Link Peer Key (TPK) handshake and Group Key handshake using WNM sleep mode frames. These KRACKs affect mobile device's roaming facilities and wireless direct connectivity features of smart TVs.

In Table 4, we examine different key reinstallation vulnerabilities (exploited using MC-MitM attacks) reported in 802.11 along with assigned CVE (Common Vulnerabilities and Exposures) identifiers from (NIST, 2021) and essential characteristics based on Common Vulnerabilities Scoring System. We highlight that leading vendor like Cisco and Google have assigned these vulnerabilities with a "high" score as millions of their Wi-Fi devices are highly affected.

Recently in 2019, (Epia et al., 2019) has recreated the KRACK on Android devices and abused all-zero-encryption key vulnerability and traced user´s private credentials from HTTP/S traffic. Fortunately, security patches are available for this vulnerability from WFA (Wi-Fi Alliance, 2017b).

Despite some KRACKs discussed in the previous section, in their follow-up paper, (Vanhoef & Piessens, 2018) presented several extensions of original KRACKs that are performed using the improved variant. In this paper, they mainly audited several available patches from WFA and some vendors and found that some are flawed and allow attacks in some instances. They also demonstrated an easier KRACK against a 4-way handshake by retransmitting an encrypted message 3 by abusing an AP's power-save functions, enabling them to attack unpatched Android devices. Most importantly, they showed a set of new key reinstallation techniques on 4-way and group-key handshake mechanism to bypass the WFA's official KRACK countermeasures by replaying the WNM (Wireless Network Management) sleep mode frames. These new KRACKs result in the encryption of data frames using an old session key so that the attacker can trivially decrypt the Wi-Fi traffic. The bypassing ability is significant- because it may enable the attacker to target even patched devices. However, the WFA has again released patches against bypassing vulnerabilities. Finally, some implementation-specific vulnerabilities are found in already patched Apple (macOS High Sierra 10.13.2) platforms that reuse station nonce values, enabling replaying handshake messages. On the other hand, Apple has patched this vulnerability.

**Table 4- Impact analysis of key reinstallation vulnerabilities**

| Assigned CVE | Handshake Details | | | Common Vulnerabilities and Exposures (CVSS VERSION 3.0) | | | | | | | Third Party Score | |
|---|---|---|---|---|---|---|---|---|---|---|---|---|
| | Type | Reinstall key | Attacker can retransmit | Base Score | Attack Vector | Attack Complexity | Privileges Required | User Interaction | Confidential-ity Index | Integrity Index | Cisco[1] | Google[2] |
| 2017-13077 | 4-Way | PTK | Message 3 | 6.8 Medium | Adjacent | High | None | None | High | High | High | High |
| 2017-13078 | 4-Way | GTK | Message 3 | 6.8 Medium | Adjacent | High | None | None | High | High | High | High |
| 2017-13079 | 4-Way | IGTK | Message 3 | 6.8 Medium | Adjacent | High | None | None | None | High | High | High |
| 2017-13080 | Group Key | GTK | Grp. Msg. 1 | 6.8 Medium | Adjacent | High | None | None | None | High | High | High |
| 2017-13081 | Group Key | IGTK | Grp. Msg. 1 | 6.8 Medium | Adjacent | High | None | None | None | High | High | High |
| 2017-13084 | Peer Key | STK | Peer Msg. 2 | 6.8 Medium | Adjacent | High | None | None | High | High | High | High |
| 2017-13087 | Group Key | GTK | WNM Msgs | 6.8 Medium | Adjacent | High | None | None | None | High | High | High |
| 2017-13088 | Group Key | IGTK | WNM Msgs | 6.8 Medium | Adjacent | High | None | None | None | High | High | High |
| 2017-13086 | Peer Key | TPK | Peer Msg. 2 | 6.8 Medium | Adjacent | High | None | None | High | High | High | High |

### *4.2. Multi-Channel MitM attacks powered by improved variant*

In (Vanhoef, 2017a), the author presented a serious implementation vulnerability in Android, Linux, and Chromium platforms, which could be effectively exploited using the improved variant. This vulnerability makes a Wi-Fi client install an all-zero-encryption key (encrypt frames with zero encryption key) instead of an actual encryption key during a 4-way handshake and enables the attacker to decrypt sensitive information effortlessly because of the absence of proper encryption during the data transmission. Nearly all implementations of Linux and Android 6.0+ platforms integrated with wpa_supplicant (v2.4 or above) are affected by this vulnerability and is exceptionally devastating against IoT devices, as most of them work on different flavors of Android or Linux platforms that internally use an affected version of wpa_supplicant. Another vulnerability enables an adversary to trick the Android clients (Chromium OS) into installing an already in-use group key. The attacker abuses the group key handshake to accomplish this task while distributing new group keys. This vulnerability critically affects most Wi-Fi devices as it enables an attacker to replay broadcast or multicast messages.

According to (Vanhoef et al., 2018), the attacker in a WPA2 network can use the MC-MitM position several ways.

1. SA query suppression can be performed whereby the MitM attacker can bypass the SA query mechanism when PMF is enabled. More specifically, after acquiring the MitM position, the attacker injects spoofed association or reassociation frames on behalf of an already connected client, which will trigger an SA query request from the AP. Consequently, the client sends back SA query responses to AP, but the MitM attacker instantly jams those responses. This causes resetting the connection (deletion of a current security association) at the AP side. The resetting of the connection makes AP unable to decrypt or recognize any packets from the client. Due to reset, the AP sends a deauthentication frame without any key (unprotected), which the PMF client would also ignore. Hereafter, the client enters into a deadlock situation as there is no way left for the client to reconnect with the AP. This attack can result in a stealthy DoS attack on PMF clients and can be possible in WPA3.

[1] https://tools.cisco.com/security/center/content/CiscoSecurityAdvisory/cisco-sa-20171016-wpa,
[2] https://source.android.com/security/bulletin/2017-11-01

2. While copying beacons, probe responses, and association frames, the MitM attacker can manipulate advertised capability and RSSI fields to deceive the clients.
3. The MSDU (MAC Service Data Unit) can be manipulated to decrypt specific MAC level data. The last two attacks happen because respective fields in beacons are not cryptographically authenticated, which is also true for WPA3.

In their work, (Chi et al., 2020) show how MC-MitM attacks could be applied in real-world settings by attacking CBTC (Communication Based Train Control) systems. The CBTC network consists of several onboard WLAN-enabled controllers that exchange sensitive train control signals over protected Wi-Fi traffic to ensure safe and efficient trains' operation. Here in CBTC systems, the attacker is connected to WLAN. That is, the attacker is an insider (as he knows the legitimate Wi-Fi passphrase) and acquires MitM position to orchestrate operations such as block, delay, inject 802.11 frames. Most significantly, since the attacker knows the Wi-Fi passphrase, he could capture Wi-Fi frames between two legitimate devices and then directly decrypt, modify, and retransmit those frames on-the-fly. To decrypt frames, the attacker derives the session key PTK by using a passphrase (PMK), MAC address of end devices, and station nonces (Anonce and Snonce) as he gets all values except station nonces. However, he traces station nonces from a subsequent 4-way handshake forced by sending forged dissociation frames to the victim. In this way, an insider MC-MitM attacker hijacks other device's communication. This attack shows the power of the MC-MitM attacker if he has the Wi-Fi passphrase in WPA2 networks. Finally, all these attacks result in redundant traction, service collapse, including emergency braking of trains. On the other hand, these attacks are not possible in WPA3 because PMK is independent of the Wi-Fi passphrase.

In the middle of May 2021, Vanhoef discovered new design flaws related to aggregation and fragmentation features in the 802.11 standards that affect every Wi-Fi device, including the devices supporting WPA3 (Vanhoef, 2021a). These vulnerabilities can be exploited using attacks dubbed as FragAttacks, which use the MC-MitM position to inject malicious packets and then obtain sensitive data from a protected Wi-Fi communication. Aggregation attacks, fragmentation mixed-key attacks, and fragmentation cache attacks are three major attacks exploiting the new design flaws in the standards (recall Section 3.5 where we illustrate the working of these attacks). In essence, the FragAttacks can be used to inject intentional packets and trick the victim into using a fake DNS server, intercept and obtain sensitive Wi-Fi communication, grab web browser cookies, or facilitate DoS attacks towards connected clients. Affected platforms or devices include, but are not limited to, macOS, Android, Linux, Windows, IoT devices, professional and home APs. Though the design flaws are serious, abusing them is not trivial in practice as they rely on some preconditions such as user interaction or rekeying.

In addition to the FragAttacks exploiting the design flaws, Vanhoef has discovered several implementation vulnerabilities due to the common programming mistakes on Wi-Fi devices. Some of them can be trivially exploited in combination with the design flaws and can be summarized as:

1. Wi-Fi devices do not verify whether fragments of the same frame possess consecutive packet numbers. The attacker can abuse this vulnerability to mix fragments from different sources through fragmentation mixed key attacks.
2. Wi-Fi devices process mixed plaintext (unencrypted) and encrypted fragments instead of accepting only encrypted fragments. This flaw allows the attacker to replace or inject plaintext instead of encrypted ones by launching aggregation or fragmentation cache attacks.
3. Wi-Fi devices forward plaintext EAPOL handshake frames to other clients even when the devices are not authenticated with the AP. This is a widespread implementation flaw found in home APs (e.g., Asus and Linksys). The attacker can abuse this flaw to perform an aggregation attack or fragmentation cache attack.
4. Wi-Fi devices that support TKIP do not check the authenticity of resembled frames. This enables the attacker to trigger fragmentation attacks to inject and likely decrypt the frames.

In Table 5, we assemble the aggregation or fragmentation vulnerabilities (exploited using the MC-MitM) with assigned CVE from (NIST, 2021). Since the FragAttacks affect almost every Wi-Fi device, WFA has released concerned patches. We congregate different MC-MitM attacks performed using the base and improved variants in Tables 6 and 7.

**Table 5- Impact analysis of aggregation and fragmentation vulnerabilities**

| Assigned CVE | Attacker can perform | Common Vulnerabilities and Exposures (CVSS VERSION 3.0) | | | | | | | Third Party Score | | |
|---|---|---|---|---|---|---|---|---|---|---|---|
| | | Base Score | Attack Vector | Attack Complexity | Privileges Required | User Interaction | Confidentiality Index | Integrity Index | Cisco[1] | Aruba[2] | Synology[3] |
| 2020-24588 | Aggregation attack | 3.5 Low | Adjacent | Low | None | Required | None | Low | Overall score is Medium, Individual score not available | Overall score is Medium, Individual score not available | Moderate |
| 2020-24587 | Fragmentation mixed key attack | 2.6 Low | Adjacent | High | None | Required | Low | None | | | Moderate |
| 2020-24586 | Fragmentation cache attack | 3.5 Low | Adjacent | Low | None | Required | Low | None | | | Moderate |
| 2020-26146 | Frag. mixed key/cache attack | 5.3 Medium | Adjacent | High | None | None | None | High | | | Moderate |
| 2020-26147 | Agg. /Frag. attack | 5.4 Medium | Adjacent | High | None | Required | Low | High | | | Moderate |
| 2020-26139 | Agg. /Frag. attack | 5.3 Medium | Adjacent | High | None | None | None | High | | | Low |
| 2020-26141 | Frag. mixed key/cache attack | 6.5 Medium | Adjacent | Low | None | None | None | Low | | | Moderate |

[1] https://tools.cisco.com/security/center/content/CiscoSecurityAdvisory/cisco-sa-wifi-faf-22epcEWu,
[2] https://www.arubanetworks.com/support-services/security-bulletins/#cat=3,
[3] https://www.synology.com/tr-tr/security/advisory/Synology_SA_21_20

**Table 6- Review of MC-MitM attacks (basic variant) in 802.11 networks**

| Ref | Security protocol affected | Attack category | Purpose of MitM | Vulnerability exploited | Attack impacts | Attack on WPA2-PMF | Affected devices/platforms | Countermeasures from authors | Patch´s availability | Possible in WPA3? |
|---|---|---|---|---|---|---|---|---|---|---|
| (Vanhoef & Piessens, 2014) | WPA TKIP | DoS attack & Break encryption | To block MIC failure reports from clients and collect packets for processing. | Flawed MIC algorithm of TKIP | Inject and decrypt wireless broadcast traffic if WPA-TKIP is chosen. | Not Applicable | All Wi-Fi devices with WPA-TKIP. | AP should initiate TKIP countermeasures fastly. | Patches are not available as WFA deprecated TKIP. CCMP can be used instead. | No |
| (Goethem et al., 2016) | WPA/WPA2 TKIP/ AES-CCMP | DoS attack | To block and forward wireless packets. | Padding is not added while encrypting MAC layer frames. | Reveal size of wireless frames, especially TCP packets and learn websites visited. | Not Applicable | Higher layer protocols such as TLS/HTTPS. | Virtual padding to avoid size information. | Not Available | Yes |
| (Vanhoef & Piessens, 2016) | WPA/WPA2 TKIP/ AES-CCMP | Downgrade Attack on 4-way handshake | To forge and inject beacons supporting only TKIP and forward messages. | AP accepts WPA-TKIP, Design flaws in the random number. | Decrypt of specific Internet traffic in a WLAN. | Not Applicable | MediaTek (flawed RNG) Broadcom (depends on OS) | APs must disable support for TKIP | Patches are not available. CCMP can be used instead. | No |
| (Vanhoef et al., 2017) | WPA/WPA2 TKIP/ | Downgrade Attack on 4-way handshake | To advertise forged beacons supporting only TKIP and inject and forward messages. | APs accept TKIP cipher suite requests when it supports both TKIP and CCMP. | Decrypt wireless traffic exploiting known vulnerabilities of RC4. | Not Applicable | All Wi-Fi devices that use Wi-Fi chip from Mediatek, Telenet, Broadcom. | RSNE parameters must be correctly verified. | Patches are not available. CCMP can be used instead. | No |
| (Vanhoef & Piessens, 2017) | WPA/WPA2 TKIP/ AES-CCMP/ AES-GCMP | KRACK on 4-way handshake | To block message 4 collect, replay and message 3. | Wi-Fi devices reinstall old PTK due to resetting of nonce and/or replay counters. | Acquire sensitive information (e.g., passwords, chats, emails), hijack HTTPs, and inject malware. | Not Applicable | All Wi-Fi capable devices are affected. Found on Mediatek, macOS Sierra 10.12, wpa_supplicant v2.3- 2.5 | Devices must verify whether the generated session key is installed once, or under one session key, the nonce or replay counter is not reused. | WFA has released official patches. (Wi-Fi Alliance, 2017a) | No |
| | | KRACK on 4-way handshake | To block message 4 collect, replay message 3. | Wi-Fi devices reinstall old GTK and IGTK due to resetting of nonce and/or replay counters. | Replay unicast, broadcast, and multicast frames. Impact IoT devices by replay of control commands. | Possible after acquiring the MitM | All Wi-Fi capable devices with Mediatek, macOS Sierra 10.12, wpa_supplicant v2.3-2.5, OpenBSD 6.1. | | | |
| | | KRACK on Group-key handshake | To block message 2, collect, replay retransmitted messages. | Wi-Fi devices reinstall old GTK and IGTK due to resetting the replay counter. | Replay group messages between the AP and the client. Hijack IoT devices while broadcasting UDP commands. | Possible after acquiring the MitM | Mediatek, macOS Sierra 10.12, iOS 10.3.1, wpa_supplicant v2.3, 2.4, 2.5 and 2.6, Windows 10. | | | |
| (Vanhoef & Piessens, 2018) | WPA/WPA2 TKIP/ AES-CCMP/ AES-GCMP | KRACK on TPK handshake | To collect, block, and replay PeerMessages | The 802.11z standard does not maintain a state machine of TPK handshake. Clients reuse nonces. | Decrypt and forge frames from smart TVs, IoT devices, and mobile phones Acquire personal sensitive information. | Not Applicable | All WPA2 devices that use wpa_supplicant versions 2.0 to 2.5. | After the first peer key message, clients shall install keys and not accept any messages after peer key message 3. | WFA has released official patches. (Wi-Fi Alliance, 2017a) | No |
| | | KRACK on Group-key handshake | To block, collect, replay WNM-Sleep Mode response messages. | WNM clients reset the replay counter while reinstalling keys. | Replay WNM Sleep Mode frames. | Possible after acquiring the MitM | All Wi-Fi devices that support WNM Mode, macOS, iOS, and wpa_supplicant version 2.6. | APs shall follow the latest IGTK in EAPOL before entering WNM sleep mode frames. | | |

**Table 7- Review of MC-MitM attacks (improved variant) in 802.11 networks**

| Ref | Security protocol affected | Attack category | Purpose of MitM | Vulnerability exploited | Attack impacts | Attack on WPA2-PMF | Affected devices/platforms | Countermeasures from authors | Patch´s availability | Possible in WPA3? |
|---|---|---|---|---|---|---|---|---|---|---|
| (Vanhoef, 2017a) | WPA/WPA2 TKIP/ AES-CCMP/ AES-GCMP | KRACK on 4-way handshake | To collect and retransmit message 3 multiple times to extend KRACK | Wi-Fi devices reinstall an all-zero session key | Decrypt client traffic from Android, Linux, and IoT devices. | Not Applicable | All Wi-Fi devices with Android 6.0 and above. wpa_supplicant v2.3-2.6, Chromium OS. | Wi-Fi chips must clear key in memory | WFA has released official patches | No |
| (Vanhoef & Piessens, 2018) | WPA/WPA2 TKIP/ AES-CCMP/ AES-GCMP | KRACK on 4-way handshake | To block and collect message 4, inject forged sleep frames to the AP, and replay message 4. | Improper power-save management in APs. | Trigger KRACK at clients. | Not Applicable | All home routers (e.g., Cisco, Aerohive, Aruba, Ubiquity) with hostapd version 2.6, Linux, OpenBSD. | Devices shall track the replay counters. Integrity of power-save frames must be verified. Clients shall store a recent GTK & IGTK | WFA has released official patches. | No |
| | | KRACK on Group-key handshake | To block, collect the first two message 3 and forward to the client after WNM frames. | Wi-Fi devices reinstall an old GTK/IGTK. | Bypass WFA´s´ KRACK countermeasure. | Possible after acquiring the MitM | | | Wi-Fi Alliance has updated the standard (Dan Harkins and Jouni Malinen, 2017). | |
| | | KRACK on 4-way handshake | To block and collect WNM-frames and broadcast frames from AP and retransmit them to the client. | Wi-Fi clients do not IGTK before going sleep mode. | Control Wi-Fi devices maliciously. Bypasses WFA´s´ KRACK countermeasure. | Possible after acquiring the MitM | | Clients shall store a recent GTK & IGTK | | |
| (Vanhoef et al., 2018) | Any | DoS on SA query procedure | To block SA-Query procedure from PMF enabled clients and send reassociation request to AP | PMF standard does not protect pre-authenticated management frames | PMF-enabled clients lose their connection from the AP. | Possible after acquiring the MitM | All PMF enabled Wi-Fi clients | Beacon protection (Vanhoef et al., 2020) may be used | Not Available | Yes |
| (Chi et al., 2020) | WPA/WPA2 TKIP/ AES-CCMP/ | DoS on CBTC systems (train control) | To collect, modify, and inject 802.11 frames between CBTC control systems. | Synchronization issues | Delayed or wrong train control, uncontrolled traction and service braking, interruption in train control, collision of two train bogies. | Not Applicable | WPA2 IoT sensors in CBTC | Not available | Not Available | No |
| (Epia et al., 2019) | WPA/WPA2 TKIP/ AES-CCMP/ | 4-Way (KRACK) | To block, replay, and forge specific wireless frames to perform all-zero key reinstallation attacks. | Wi-Fi devices reinstall an all-zero session key | Recover the user details (e.g., username and password) when the victim visits certain websites using Android devices. | Not Applicable | Android 7.0 or above | Wi-Fi chips must clear key memory. | WFA has released official patches. | No |
| (Vanhoef, 2021a) | WPA/WPA2/ WPA3/TKIP/ CCMP/ GCMP | Frame aggregation attack | To flip IPv4 packet into aggregated (A-MSDU) frame, | Aggregation flag in the frame header is not authenticated | Inject arbitrary packets, trick the client towards fake websites, mix malicious fragments, obtain or decrypt user´s sensitive data. | Applicable | All Wi-Fi devices, Linux, Windows, macOS, iOS, IoT devices, routers (Cisco, Aruba, D -Link), NICs. | Ensure A-MSDU flag is authenticated in all frames. | WFA has released official patches. (Wi-Fi Alliance, 2021) | Yes. Attack reported on devices including WPA3. |
| | | Frame fragmenta-tion attack | To intercept, block, or forward specific fragments. | Lack of verification of fragments sent by different users, fragment cache not cleared. | | | | Fragments encrypted by different keys must not be processed, Cache must be cleared when (re)connection occurs. | | |

### 4.3. Challenges in the adoption of general protection mechanisms

In this subsection, we discuss the significant challenges in adopting security patches (against KRACK and FragAttacks) and PMF in reducing the impact of MC-MitM attacks.

#### 4.3.1. Challenges in security patching

As it can be observed from Tables 6 and 7, the MC-MitM position was widely used to trigger attacks like security downgrade attacks, DoS attacks, implementation-specific exploits, KRACK, and including the latest FragAttacks towards the protected Wi-Fi traffic. Fig.15 shows the statistics of analyzed MC-MitM enabled attacks.

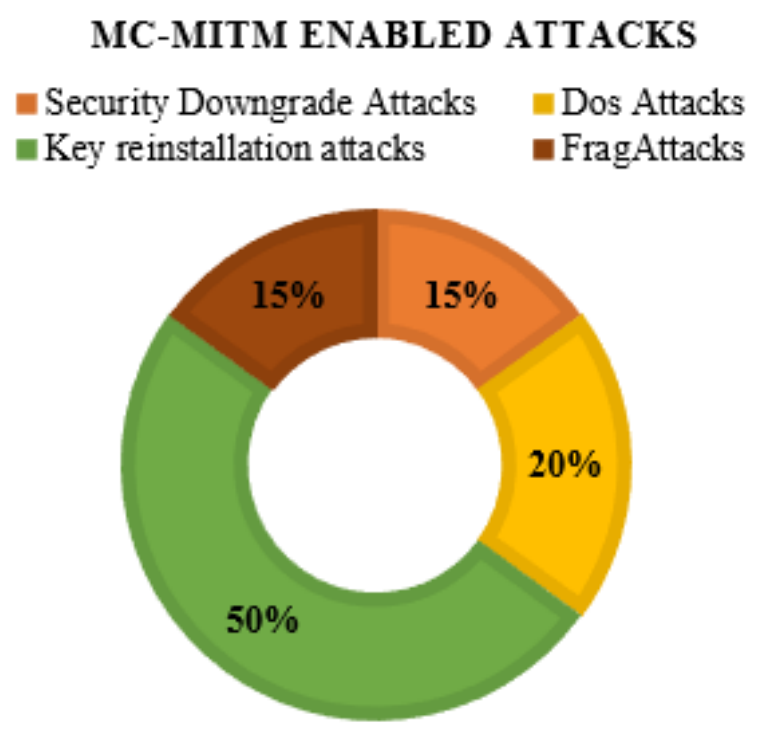

**Fig. 15. Statistics of MC-MitM enabled attacks.**

Amidst different attacks, the key reinstallation attacks and FragAttacks are most significant, which provide multiple ways to launch MC-MitM attacks due to the critical design flaws in the core handshake mechanisms and aggregation or fragmentation features of 802.11 standard. Since these attacks exploit flaws in the 802.11 standards, there is a high risk to every Wi-Fi device if the respective vendors have correctly implemented those standards. On the other hand, the WFA and affected vendors have released corresponding patches to prevent KRACK or FragAttacks. However, the available patches can only be applied to robust wireless clients (e.g., desktops, laptops, smartphones, professional routers) with provision for managing software or firmware patches in a much more hassle-free manner. Affected devices include millions of Wi-Fi devices connected to the Internet of Things (IoT) networks. Patching security vulnerabilities of key reinstallation, aggregation, or fragmentation can be challenging for several reasons, as discussed below.

***4.3.1.1. Lack of Security Patches.*** IoT devices might most likely miss security patches against KRACK or FragAttacks due to insufficient patch support from respective vendors or companies. This is mainly because IoT companies release their devices, delivering seamless and hassle-free connectivity services at minimum cost, and adding continuous support increases the costs of deployment and maintenance. Additionally, to apply key reinstallation patches successfully, an IoT device requires an update of its underlying firmware and patches from the affected chip vendors (chip partners) that must be applied on devices' firmware patches (WILBUR, 2017). This requirement brings massive responsibility for device vendors because they must first get updates from corresponding chip partners to release their new firmware patches. The conundrum is that device vendors do not release their patches because of limited update support periods even though chip vendors release their patches, while the reverse scenario is also possible.

Furthermore, IoT companies always go for dynamic changes for incorporating new services in their device to grasp the fast-paced growth of the Internet of Things markets. Thus, updates may not be available to devices as they neglect older devices or those devices with no sufficient market profit. Of great concern is that often vendors do not release patches even if responsible authorities notified them. For example, according to the CERT coordination center's vendor details shown in (CERT, 2017), we can see that only 17% of notified device vendors have released patches during the coordinated patch release period during 2017. Fig.16 shows those statistics from the vendor information page of CERT. Generally, well-known vendors such as Google, Microsoft, Apple, and other famous router manufacturers have released patches. However, patch release details of many vendors, including IoT manufacturers, are unavailable as per CERT.

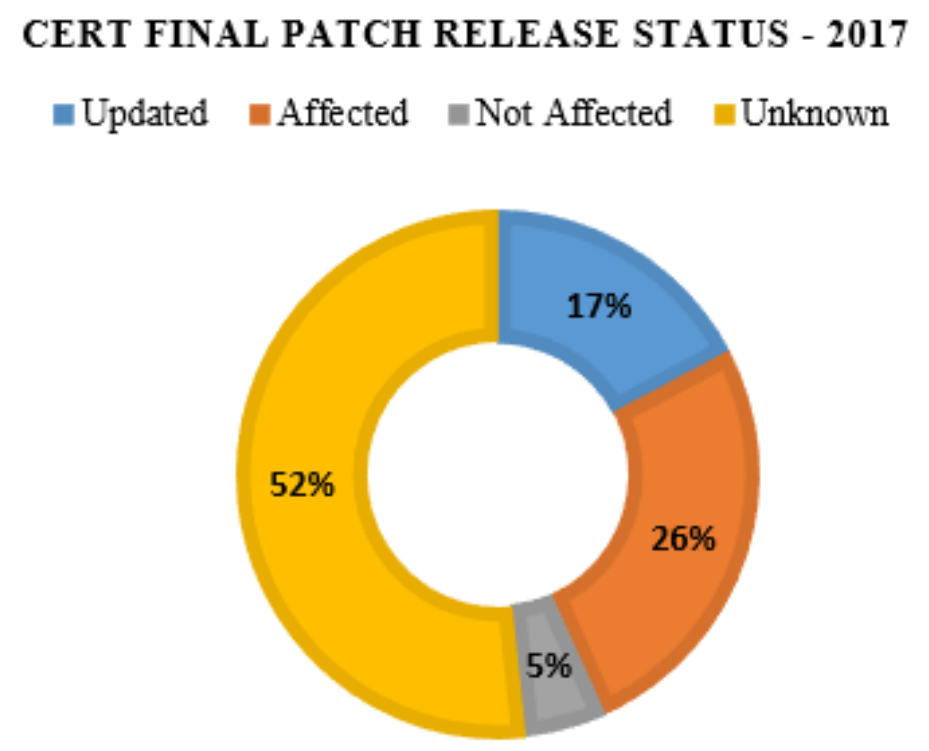

**Fig. 16. Statistics of KRACK patch release by CERT.**

Similarly, as shown in Fig.17, the company Security Focus widely tested Wi-Fi devices (e.g., servers, operating systems, routers, Wi-Fi chips, IoT devices) in the year 2019 (Security Focus, 2019). They reported that 90% of tested devices, including, but not limited to devices from Cisco, Aruba, Google, Microsoft, Intel, Apple, and Siemens, are vulnerable to key reinstallation attacks. Nearly all devices are affected here because, even if vendors release their new products or implementations, they generally ignore key reinstallation patches or test their new implementations against such vulnerabilities before commercializing them.

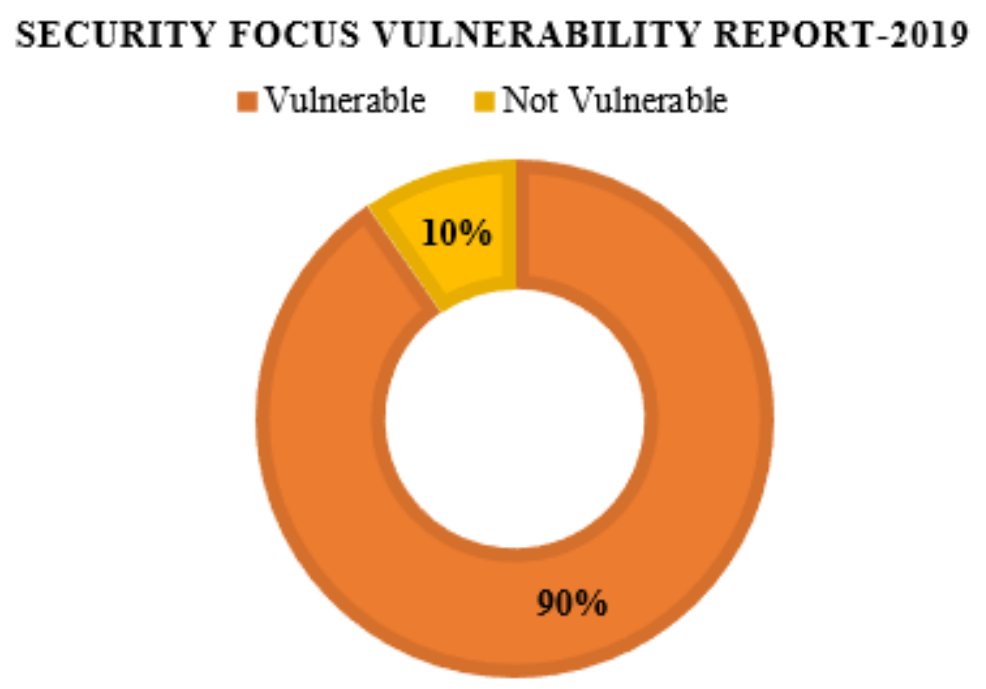

**Fig. 17. Statistics of KRACK patch release by Security Focus.**

A recent experimental study (August 2020) presented by (Freudenreich et al., 2020) also concludes that several mobiles and IoT devices (around 65% among tested) are still vulnerable to different types of key reinstallation attacks due to the unavailability of patches from respective device manufacturers.

All these statistics show that patching key reinstallation vulnerabilities or defending against KRACK is still a considerable dilemma even 4 years after their discovery. The main reason is, in reality, not every vendor releases patches responsibly for their new and old devices. So, it can be assumed that the same patching problems will continue for several years with the new FragAttacks. In other words, negligence in releasing patches makes most of the Wi-Fi devices in our home networks continue unpatched or exposed to MC-MitM enabled attacks.

On the other hand, although patches are available for MC-MitM enabled attacks, especially for KRACK and FragAttacks for WPA2 or WPA3 devices, there are no security patches for certain vulnerabilities related to WPA devices (e.g., (Vanhoef & Piessens, 2014), (Vanhoef et al., 2017), (Vanhoef & Piessens, 2016)). Even the available KRACK or FragAttacks security patches may not be applied on WPA devices because Wi-Fi Alliance deprecated TKIP in 2015 (Wi-Fi Alliance, 2015). WPA devices may be patched if the respective vendor specially develops patches for the vulnerabilities, but such actions are uncommon in practice. Unfortunately, several existing legacy Wi-Fi devices (e.g., smart TV, smart refrigerator, smart bulb), and mostly constrained IoT devices, are still working on TKIP to comply with their low-computing resources. Similarly, routers used in our home or office settings are still operating on TKIP/CCMP transition mode to avoid interoperability issues. The use of PMF is also not possible on WPA devices. A recent (Sept 2020) survey on Wi-Fi network issues conducted in (Reyes et al., 2020) critically shows that more than 50% of analyzed devices employ WPA-TKIP. This is a critical condition where MC-MitM attackers will have multiple opportunities to potentially inflict damage on Wi-Fi environments by targeting those WPA-TKIP devices.

***4.3.1.2. Patching Difficulties.*** IoT device's realm experiences an enormous difficulty in dealing with updates, mainly due to applicability of patches on them. In many situations, IoT devices arrive with static programming or non-upgradable firmware models. This prevents such proprietary IoT devices from subsequent user-serviceable upgrading of the device or the Wi-Fi chip used in it (Chin & Xiong, 2018). Similarly, the security patches may not always comply with IoT device's firmware due to a mismatched vendor model, model of Wi-Fi chips, versions of hardware, or underlying operating systems. Fixing key reinstallation vulnerability is also risky because it likely damages the firmware of IoT devices. Another issue is the lack of I/O capabilities. For example, smart refrigerators, smart locks, window blinds, etc., often have no easily accessible user interfaces, and thus applying patches on them is difficult. Users also find difficulty in downloading patches as many IoT devices may not support over-the-air (OTA) updates (Lin & Bergmann, 2016). Additionally, to effectively defend against key reinstallation attacks, every device connected to the Wi-Fi network must be appropriately patched. Most clearly, every client and AP must be applied with patches, which is not usually feasible, especially when there are several heterogeneous devices in WLAN or home IoT settings. Updating only the affected router or client is not sufficient because even one unpatched device on a network can become a vulnerable component for MC-MitM attackers. Moreover, KRACK or FragAttack vulnerabilities have a set of more than ten security patches (Wi-Fi Alliance, 2017b, 2021) that must be applied separately on target devices. This makes the patching process more challenging, and thus, holistic patching is not practical, especially in IoT networks.

***4.3.1.3. Lack of Technical Knowledge.*** While most people are aware of the key reinstallation vulnerabilities, they struggle or sometimes never perform patching due to the lack of substantial technical knowledge (Freudenreich et al., 2020). Sufficient device handling and installation skills are required for patching security flaws and bugs on Wi-Fi-capable devices. For devices like smartphones, this task is easy as it provides automatic push notifications and requires permission from the users. Similarly, if the vendor adequately maintains an IoT device by releasing patches, the user can apply the firmware patches through the connected mobile application. Patching some IoT devices (e.g., Raspberry Pi) is also difficult for common people as they need to download firmware according to the kernel version and then use specific Linux commands to apply the firmware patches. To apply patches on the APs, the user has to download firmware images of their router using its model number and firmware version. Then through the router web interface, he has to apply firmware upgrade by selecting the corresponding firmware images if the router does not provide automatic firmware update provisions. Additionally, users must be aware of rollback procedures in case of any firmware failures. In all cases, a substantial amount of technical knowledge is required.

***4.3.2. Challenges in adopting PMF***

Generally, PMF is used to defend against DoS attacks like deauthentication or disassociation attacks as part of MitM attacks from outsiders. Although PMF can resist these attacks, its adoption in existing WPA2 networks is quite challenging due to the following difficulties:

- PMF can defend DoS or MitM attacks only if every AP and client in a Wi-Fi network supports it. A PMF capable AP cannot admit a client that does not support PMF and vice versa. In personal Wi-Fi networks, the AP rarely supports the PMF and is available mostly if APs support 802.11n or 802.11ac standards. Generally, only high-end routers (e.g., Cisco) support PMF in enterprise networks (Cisco, 2020).
- It is generally difficult to enable PMF on existing Wi-Fi or IoT devices because proper software or firmware upgrade is required not just for an AP but also for every client (Cisco, 2017; CWNP, 2009). On the other hand, it is not possible to enable PMF unless device vendors support it.
- When PMF is enabled, some devices connect to the network for a short time and may suddenly get disconnected. On some devices, enabling PMF does not show an IP. Certain Wi-Fi clients do not support PMF if it runs on Wi-Fi version 4 or below (Cisco, 2020; Telstra Air, 2020).
- PMF may create many compatibility issues as it requires support from both the operating system (OS) and the Wi-Fi chip's driver used in devices (Cisco, 2017). For example, If OS supports, the chip's firmware may not always support 802.11w, or there will be no patches available for specific devices. It's generally unknown the devices or firmware versions that come with PMF.
- PMF cannot protect legacy Wi-Fi devices operating on TKIP to cope with their low-computing resources (CWNP, 2009; Reyes et al., 2020). Additionally, software patches cannot be applied on such devices as the WFA deprecated WPA-TKIP.
- PMF standard itself is vulnerable to key reinstallation attacks (CVE-2017-13081). Therefore, difficulties of KRACK security patching discussed in the previous section will also affect its use in real-world Wi-Fi applications.

On the other hand, PMF cannot defend against DoS attacks based on Wi-Fi jamming as well as rogue AP-based threats by spoofing the beacons (CWNP, 2009). This allows especially MC-MitM attackers to deceive WPA2 or WPA3 devices, even if PMF is enabled. Additionally, an insider MitM attacker can trigger deauthentication, disassociation attacks as he is authorized to access the network and so is the case with MC-MitM attackers.

### *4.4. MC-MitM attack scenarios in WPA3 networks and possible impacts*

In this subsection, we analyze the possible impacts of MC-MitM attacks in WPA3 networks because of their ability to circumvent PMF protection. We create relevant attack scenarios where MC-MitM attackers can pose critical challenges in WPA3 networks.

#### *4.4.1 Connection behavior of clients in WPA3 networks*

This section depicts the connection behavior of the clients in WPA3 networks. As per Table 8, WPA3-Personal can be configured in two security modes: WPA3-Only mode and WPA3-Transition mode. In WPA3-Only mode, the AP accepts clients that support only WPA3 that use PMF by default. When WPA3-Transition mode is used, the AP accepts both WPA2 and WPA3 clients. Additionally, the AP can be set either as “required” or “enabled” modes in this configuration. In the “required” mode, the AP only accepts WPA2 or WPA3 clients with PMF, and in the “enabled” mode, the AP also accepts WPA2 clients without PMF. Important to note that WPA3 does not provide backward compatibility for WPA-TKIP clients.

**Table 8 - Client connection behavior in WPA3 networks (Cisco, 2021)**

| **Security mode** | **PMF** | **Connection behavior of the client** | | |
|---|---|---|---|---|
| | | WPA2 Client | WPA2-PMF Client | WPA3 Client |
| WPA3-Only | Required | Cannot connect | Cannot connect | Connection Possible |
| WPA3-Transition | Required | Cannot connect | Connection Possible | Connection Possible |
| | Enabled | Connection Possible | Connection Possible | Connection Possible |

#### *4.4.2 MC-MitM attack scenarios in WPA3 networks*

To deceive any device connected in a WPA3 network, the MC-MitM attacker can adopt either base or improved attack variants. Here, for the sake of analysis, we use MC-MitM improved variant attacks. However, the principal impact is the ability of MC-MitM attacks to circumvent PMF protection in acquiring the MitM position. Further, the following are the two different WPA3 attack scenarios, which amplify the impact of attacks.

***4.4.2.1. WPA3-Only mode attack scenario and impacts.*** As depicted in Fig.18, the MC-MitM attacker can target any of the WPA3 clients in the attack scenario. Once the attacker deceives a WPA3 client, he can block or modify any frames between the end devices and induce different kinds of FragAttacks (Vanhoef, 2021a). He can also perform DoS attacks such as SA query suppression and eventually disconnect the WPA3 client from the legitimate network. Size exposing attacks (Goethem et al., 2016) may also be effectively used to learn about the victim's private web traffic. Additionally, it is possible to modify advertised capabilities such as bitrates in beacons or probe response to control data bandwidth. According to (MTROI, 2014), when the attacker gains access to a WPA3 network (insider attacker), he can also send authenticated channel switch announcements through protected action frames and steer clients to connect his rogue channel. However, the MC-MitM attacker cannot perform KRACK or other kinds of offline dictionary attacks on WPA3 networks.

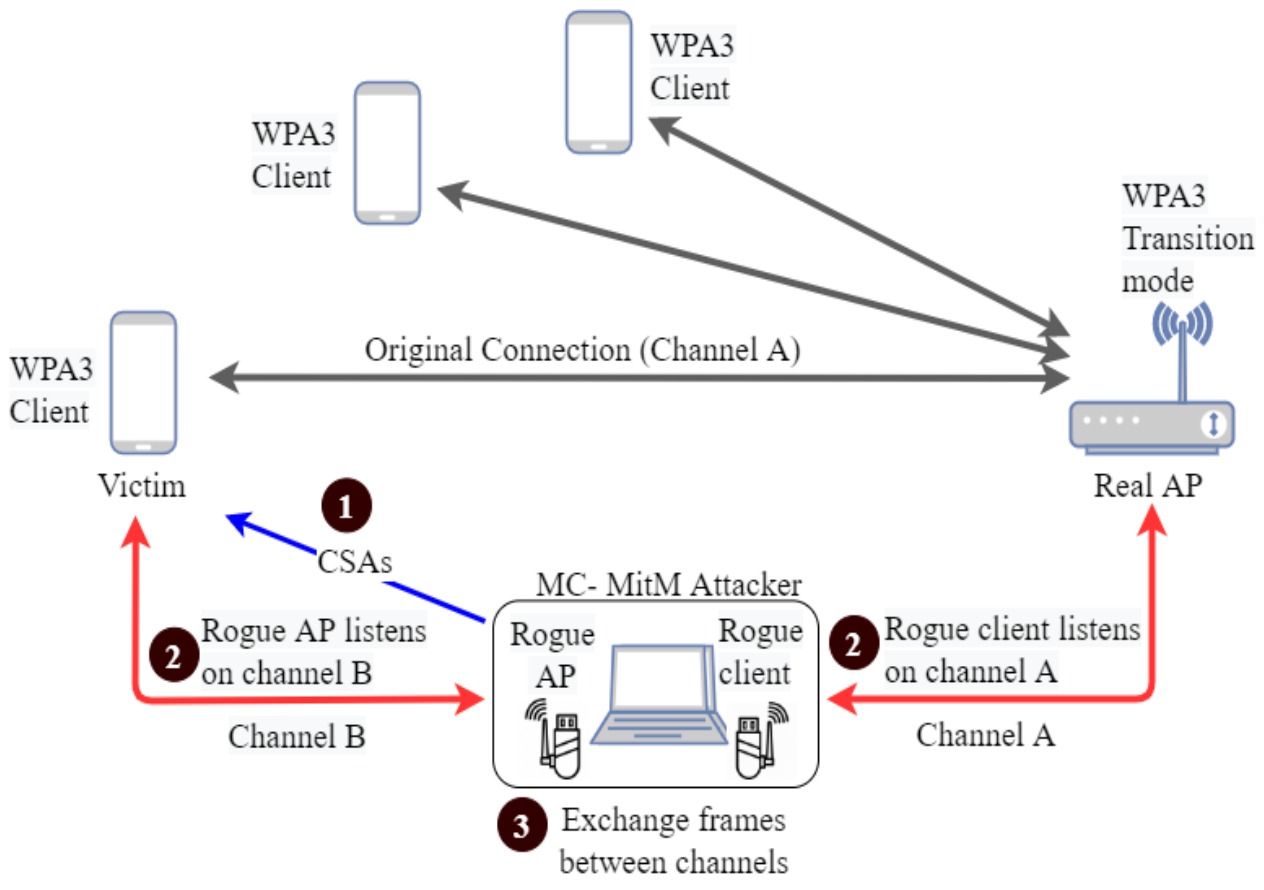

**Fig. 18. WPA3-Only mode attack scenario.**

***4.4.2.2. WPA3-Transition mode attack scenario and impacts.*** In transition modes (required and enabled) of WPA3 shown in Figs.19 & 20 respectively, both WPA2 and WPA3 clients share a common Wi-Fi passphrase. So, with these attack scenarios, the MC-MitM attacker may target a WPA2-PMF or regular WPA2 client and capture specific four-way handshake messages to perform dictionary attacks. If found, attackers can challenge WPA3 networks by retrieving the password. Attackers can also decrypt previously encrypted WPA2 sessions but not WPA3 sessions. Though these attacks do not require a MitM position, the MC-MitM would facilitate such attacks more efficiently. Furthermore, KRACK is possible on both WPA2-PMF and regular WPA2 devices. SA query suppression and FragAttacks can also be performed on any WPA2 or WPA3 devices. All these attacks can potentially challenge the security of WPA3 networks.

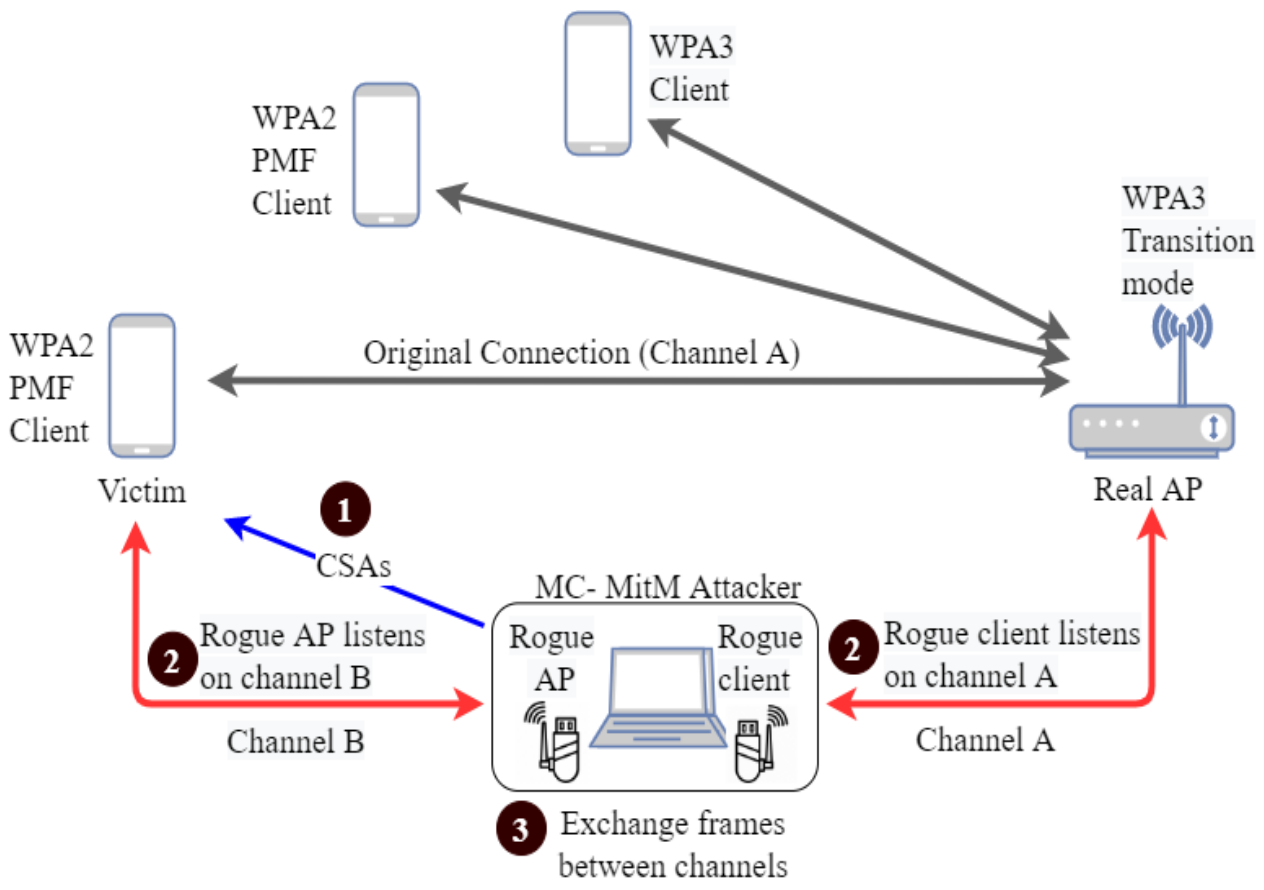

**Fig. 19. WPA3-Transition mode-required attack scenario.**

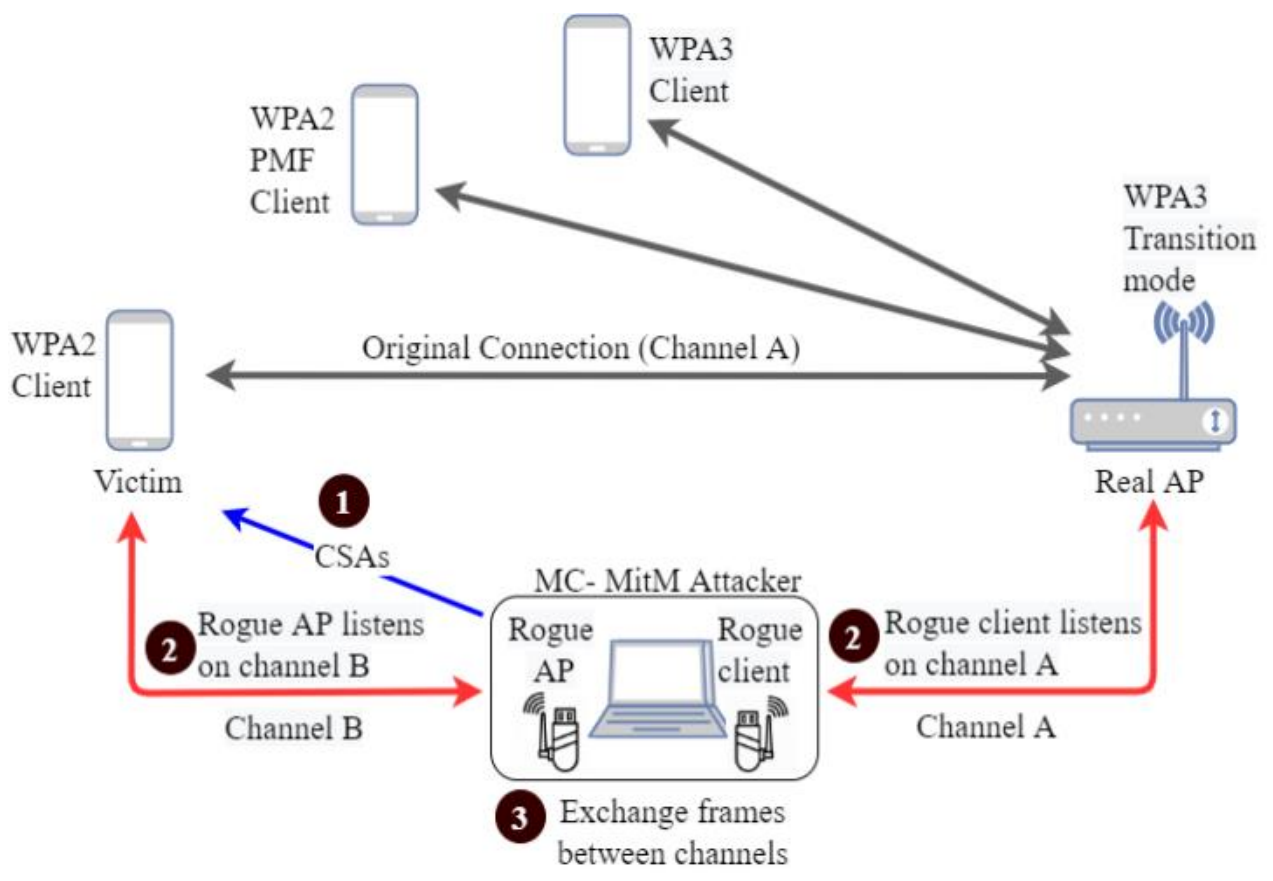

**Fig. 20. WPA3-Transition mode-enabled attack scenario.**

## 5. Multi-Channel MitM Defense Mechanisms

In this section, we analyse existing defense mechanisms for MC-MitM attacks. We also evaluate the feasibility of existing mechanisms for deploying in real-world IoT settings or environments.

### *5.1. Classifications of detection mechanisms*

As shown in Tables 4 and 5 MC-MitM attacks have been a trend in attacking protected 802.11 networks since 2014. Based on the purpose or application of the defense mechanisms, we classify them into two stages:

- **Stage 1 defense mechanisms**: This category focuses on defending against attackers before acquiring a MC-MitM position by recognizing real attack vectors, such as rogue devices, rogue channels, or spoofing channel switch announcements.

- **Stage 2 defense mechanisms**: This category focuses on defending against MC-MitM enabled attacks (e.g., KRACK, FragAttacks, DoS) or other attacks after acquiring a MitM position.

### *5.2. Analysis of stage 1 defense mechanisms*

The first stage 1 defense mechanism is Operating Channel Validation (OCV), proposed by (Vanhoef et al., 2018), which cryptographically validate the parameters defining the operating channel between two wireless stations. They introduced a new Operating Channel Information (OCI) element (as an extension to the 802.11 standards) to be included in EAPOL frames and is verified during handshake processes (e.g., 4-way, group-key). In essence, on receiving handshake messages, the receiver verifies whether the OCI is present and the primary channel used for communication matches the one in the OCI of the sender. When a mismatch occurs, the OCV aborts the current handshake and prevents the attacker from acquiring the MC-MitM position. Besides, for preventing unprotected channel switch announcements through beacons or probe responses even when PMF is enabled, the authors proposed to include OCI in the SA query request-response messages of PMF.

To whatever extent the authors point out that the OCV still allows obtaining partial MitM (the attacker will be successful only if the AP sends CSAs). Here, the attacker tracks CSAs from the AP and jams them to keep the client stay on the old channel. He also captures and stores all frames from the client. Meanwhile, the attacker sends spoofed CSAs to the client before the AP starts disconnecting clients due to the SA query timeout. This will force the client to switch channels and complete the SA query. As of now, the attacker sends the previously stored frames to the AP and acquires the MitM. Similarly, it is also possible to acquire partial MitM (the attacker gains frames only from the client) by exploiting the specific bandwidth parameters that are not authenticated cryptographically. Though the impact is less, the possibility of partial MitM can allow the attacker to bypass the OCV. Additionally, insider attackers can send protected channel switch announcements through action frames and perform SA query suppression (jam specific SA query messages), causing the resetting of the client's connection or DoS.

In their work, (Vanhoef et al., 2020) have proposed a stage 1 defense mechanism known as beacon protection to defend against attacks abusing unprotected beacons. Their main aim was to prevent rogue AP-based attacks such as silencing stations, power constraints manipulation, possible partial MitM attacks in (Vanhoef et al., 2018), channel switch announcements in MC-MitM attacks, etc. They introduced an extra information element (IE) in every beacon so that the clients can verify it when connecting to an AP. To achieve this, they modify the Management Message Integrity Code Element (MME) of the Broadcast Integrity Protocol (BIP), which is a part of the PMF standard. In this mechanism, a Beacon Integrity Packet Number (BIPN) in the beacon is incremented after every transmission so as to detect spoofed or replayed beacons. Notably, a new group key called Beacon Integrity Group Temporal Key (BIGTK) will be distributed to every client when they connect and authenticate with an AP. This enables every client to generate the Message Integrity Code (MIC) to verify and authenticate beacons from legitimate AP and ignore any unauthorized ones without MME or invalid MIC value, thus avoiding the risk of rogue APs to an extent.

Beacon protection may effectively protect beacons or probe responses; however, it does not consider certain unauthenticated action frames with channel switch announcements that can be abused even if PMF is enabled. Moreover, to realize beacon protection in practice, every client needs to store a reference beacon frame before connecting to an AP to verify beacon´s legitimacy using an already distributed group key (BIGTK). This requirement may be challenging to achieve, especially with constrained IoT devices having no access control or storage capabilities. Finally, the proposed mechanism does not block insider attackers. For example, suppose the MC-MitM attacker is connected to WLAN. In such scenarios, he can still introduce MC-MitM attacks (send CSA action frames to steer clients to his rogue channel) in a much easier manner as he is authorized to perform network operations. This may result in the hijacking of private communication of other users or devices inside homes or offices. (Chi et al., 2020) is a good example of such an insider MC-MitM attack.

We highlight that the aforementioned defense mechanisms, i.e., (Vanhoef et al., 2018) and (Vanhoef et al., 2020), are incorporated in 802.11 standards, and recently, in December 2020, the WFA included them in WPA3-2020 updates as optional features (Stephen Orr, 2020). Nonetheless, the effectiveness of both mechanisms depends heavily on stringent security conditions such as the support for PMF, especially to defend spoofing of channel switch announcements (CSAs) in WPA and WPA2 networks and the need for software or firmware patches for WPA, WPA2, as well as WPA3 devices (due to changes in handshake procedures).

In WPA3-2020 updates, the WFA also included another feature known as SAE-PK (Public Key) to uniquely identify APs in a WLAN during the connection establishment process based on ECC public key cryptography (Wi-Fi Alliance, 2020 § 6). This can be considered as a stage 1 defense as it prevents insider attackers from setting up rogue AP and performing MitM attacks. To implement SAE-PK, the network administrator generates a passphrase that acts as a fingerprint of the legitimate AP with which a client can connect to protected Wi-Fi (private or public) networks. SAE-PK authentication is an extension of regular SAE with an additional confirm message from the AP to the client consisting of the digital signature of the AP´s public key. As a result, the client can verify this digital signature using the public key. Therefore, even if the attacker knows the passphrase, he does not know the corresponding private key used to generate a valid digital signature. Consequently, the insider attacker would not be able to set up rogue AP and perform MitM operations. However, we conjecture that SAE-PK will not prevent MC-MitM attacks. This is mainly because rogue APs are identified only during the SAE-PK authentication phase or when the client connects to the AP for the first time. On the other hand, the MC-MitM attacker usually acquires a MitM position between an already connected client and the AP. He can also bypass the SAE authentication because, according to (Huawei, 2020), the WPA3 client uses an open authentication instead of an SAE authentication while reconnecting to an already authenticated or connected network.

Aware of the partial MitM based attacks in (Vanhoef et al., 2018), (Chatterjee et al., 2020) defined a stage 1 defense mechanism based on Physically Unclonable Functions (PUF) to prevent rogue AP´s actions during the MC-MitM attack. The PUF is a digital fingerprint that can act as a unique identifier for an electronic circuit board structure, which is very difficult to clone since no two devices can have similar PUF based identifiers. The basic idea is to generate a unique secret key from the AP's PUF signature and use it to mutually authenticate devices (the AP and client). A dedicated server (trusted third party) stores a PUF signature (a challenge-response value pair, aka CRP) of the AP in WLAN and assigns a secret key to every client. When a client wants to join a particular AP, it communicates with the server and proves its legitimacy using a secret key. Therefore, an attacker who does not know the PUF signatures of their rogue AP will not make the authentication successful, thereby blocking key reinstallations or related MC-MitM attacks using rogue APs. However, PUF based authentication itself is under threat of several kinds of MitM attacks (Babaei & Schiele, 2019).

In yet another stage 1 defense mechanism, (Gong et al., 2020) proposed an anomaly detection system for the Wi-Fi clients to find rogue AP´s actions during the MC-MitM attack. To find anomalies during the connection establishment, they modify the source code of the wpa_supplicant (an open-source implementation for Wi-Fi clients) and install it on every Wi-Fi client in a WLAN. The modified wpa_supplicant verifies the uniqueness of a pair of BSSID (MAC address of AP) and ESSID (network name) when a client begins connecting to an AP. If they are not unique, the mechanism prevents the clients from connecting to that particular AP and alerts users. However, the effectiveness of the proposed anomaly detection depends only if the attacker uses reactive jamming that often produces less lag in receiving beacons from the legitimate AP so that the client can decide by comparing these beacons with that of a rogue AP. On the other hand, if the MC-MitM attacker uses a continuous jammer on the legitimate AP´s channel, AP´s signals or beacons will not be unavailable for the target client, making the detection difficult. Moreover, depending only on the uniqueness of the BSSID and ESSID pair will not be effective because there can be many situations with the same pairs of identities. For example, when the AP supports a dual-band connection, there can be chances to have the same pair of such identities.

Although the defense mechanisms by (Chatterjee et al., 2020) and (Gong et al., 2020) can harden MC-MitM attacks by analysing the uniqueness of the rogue AP´s identities, their practical adoption may be difficult in real-world scenarios. This is because, in the former one, PUF authentication can be implemented only on FPGA (Field Programmable Gate Array (FPGA) devices, and its extraction is impossible with proprietary or commercially available routers (Babaei & Schiele, 2019). Moreover, it requires a sophisticated software tool provided by the FPGA manufacturer for subsequent programming and configurations. In the latter one, installing wpa_supplicant may be possible on embedded devices (e.g., Raspberry Pi), but the installation can be challenging on proprietary Wi-Fi devices that use specific software/hardware from the vendors.

### *5.3. Analysis of stage 2 defense mechanisms*

As soon as the reinstallation attacks reported in 2017, (Chin & Xiong, 2018) introduced a stage 2 defense mechanism known as KRACK-Cover in, which helps Wi-Fi end-users to detect the presence of key reinstallation attacks in a WLAN. The proposed mechanism first captures and analyses 802.11 MAC layer frames in the target network by using sensors followed by validating message configurations of frames. The mechanism then identifies respective packets transmitted from validated frames while executing the KRACK attack scripts, including retransmitted broadcast/multicast frames or retransmitted 4-way handshake messages targeting different clients. Finally, the system alerts the end-user with a warning message upon finding such dubious handshake messages present during executing KRACK attacks.

As a subsequent stage 2 defense to detect key reinstallation attempts, (Naitik et al., 2018) presented a detection mechanism for clients in a WLAN. Their system first collects 802.11 MAC layer frames and then extracts WPA key data from 4-way handshake frames to know nonces' value. This is followed by verifying whether duplicate message 3 (EAPOL frame) is present in the wireless network stream. The AP retransmits message 3 when the attacker blocks message 4 from the victim to the AP. Once duplicate message 3 is found, the detection mechanism generates alerts to the administrators. Closely related to (Naitik et al., 2018), Natital developed a KRACK attack detector using python scripts in (Securingsam, 2017). This script can be run on open-source APs (e.g., hostapd) rather than clients. It identifies any duplicate message 3 of the 4-way handshake in a particular WLAN and disconnects the suspected device, preventing it from sending any further sensitive data to the AP.

The defense mechanisms proposed by (Naitik et al., 2018) and (Securingsam, 2017) manage to identify retransmitted message 3 of the 4-way handshake during KRACK attacks executed using the MC-MitM position. As per the 802.11 standards, it is quite reasonable that an AP retransmits message 3 in many circumstances. For example, retransmission occurs due to network traffic congestion, or it may continue until the AP reaches its maximum retransmission limit. Therefore, blocking every retransmitted handshake message may result in frequent handshake failures. Instead, systems could have verified whether the same session key was reused in subsequent retransmissions.

In another work, (Abare & Garba, 2019) enhanced the stage-2 defense mechanism by (Naitik et al., 2018) and proposed prevention mechanisms to authenticate handshake messages against key reinstallation attacks. Here, to avoid forging WPA key data nonce values and retransmission of message 3 of the 4-way handshake, the proposed mechanism encrypts complete handshake messages, including nonce values Wi-Fi pre-shared key. While encrypting the handshake's first message (from AP to the client), they include a new Boolean value initialized to TRUE with other standard parameters. On receiving this, the client decrypts it and stores the Boolean value. The client then encrypts this Boolean value with the necessary parameters and forwards it to the AP in message 2. If the subsequent decryption is successful, the AP forwards message 3 with the client's respective MIC and otherwise, it aborts the handshake. After decrypting message 3, the client changes the Boolean value to FALSE before sending message 4 to the AP, which indicates that the pairwise key is installed once. Thus, by verifying the Boolean value, the client can detect and prevent the repeated installation of keys later when message 3 is retransmitted during key reinstallation attacks. Significant to note that this prevention mechanism mandates changes in the Wi-Fi standard.

A software-defined networking (SDN) based stage 2 defense mechanism is introduced by (Li et al., 2019) to defend key reinstallation attacks. The proposed mechanism consists of detection and prevention modules, and are hosted on the AP in a WLAN. The SDN controller parses and inspects each incoming Wi-Fi network frame to trace any duplicated message 3 of the 4-way handshake to detect attacks. Additionally, it verifies the nonce and replay counter value in the frame to ensure whether there is any key that has been reused. To prevent attacks, this mechanism requires the AP to be configured to work as an Open Flow Switch (OVS), which is a programmable network protocol for SDN environment. Once the SDN controller detects the attack, the prevention module updates attack details in the flow table's entries in the OVS and then redirects the attack traffic flows to a splash portal, a disk space to store attack traffic instead of forwarding it to the client.

Though defense mechanisms by (Abare & Garba, 2019) and (Li et al., 2019) provide detection and prevention of KRACK attacks, they focus only on basic KRACK attacks, i.e., retransmission of message 3 during a 4-way handshake. However, attackers can still instil other forms of KRACK attacks (e.g., group-key, peer-key, and WNM sleep mode frames) even with the above defense mechanisms.

(Cremers et al., 2020) have enhanced previous stage 2 defense mechanisms by developing completely new handshake protocols for preventing different forms of key reinstallation attacks. These protocols identify the nonce-reuse weaknesses of underlying cryptographic algorithms, thereby improving the security of handshake mechanisms in 802.11 standards, and are basically security patches that manage the nonce and replay counter reuses 4-way handshake, group key handshake, WNM sleep mode, etc. They also claim that their protocols can defend against key reinstallation attacks even in the absence of previous stage 1 defense mechanisms. Another formal model proposal can be presented in (Chothia & Ryan, 2020) that prevents different forms of KRACK and also defends against cipher suite downgrade attacks on APs. However, there is no evidence that these formal models are tested in real-world attack scenarios.

Traditional Intrusion Detection Systems like SNORT (Marty Roesch, 2021) released rules for detecting KRACK attacks in 2018 (SNORT, 2018). We consider SNORT as a post-attack defense mechanism since it identifies KRACK attacks triggered after acquiring the MC-MitM position. SNORT rules filter and detect malicious network packets with specific contents (e.g., Dot11, RadioTap, and FCfield) that may occur while running KRACK attack scripts. These filtering contents are key components of the KRACK python script and Scapy (a packet manipulation tool) utilities. However, the contents used by SNORT rules for detecting or matching KRACK can even be present in typical WLAN packets or scripts of other attacks developed using Scapy. Hence, employing SNORT with this specific rule may be ineffective or generate false alarms.

### *5.4. Technical feasibility analysis of MC- MitM defense mechanisms*

In this subsection, we define specific qualitative metrics to evaluate the technical feasibility of implementing stage 1 and stage 2 defense mechanisms against MC-MitM enabled attacks in real-world IoT environments. We assume that IoT environments host Wi-Fi supported constrained devices like smart lights, smart sensors (e.g., temperature, humidity, pressure), smart controllers (e.g., plugs, switches, curtain, door), smart appliances (e.g., thermostats, refrigerator, washing machine, oven) along with other robust devices such as home routers (APs), smartphones, laptops or computers.

#### *5.4.1. Metrics used for technical feasibility analysis*

We consider undermentioned metrics to evaluate the technical feasibility of existing defense mechanisms.

- **Changes in the Wi-Fi standard**: This metric indicates whether the proposed defense mechanism requires protocol changes in any of the existing Wi-Fi standards (802.11 or 802.11w).
- **Defense mechanism installation/compatibility:** This metric indicates whether the proposed defense mechanism requires the installation of new capabilities or expects their compatibility on every device (Wi-Fi client, AP) for successful implementation.
- **PMF requirements:** This metric indicates whether the proposed defense mechanism requires PMF on every device for its implementation.
- **Firmware updates:** This metric indicates whether the proposed defense mechanism requires firmware updates on every device to successfully execute new defense mechanisms or enable specific network configurations (e.g., PMF). Firmware updates are also required if the defense mechanism mandates changes in Wi-Fi standards.
- **Third-party software/hardware integration:** This metric indicates whether the proposed defense mechanism requires installing any third-party software (other than defense mechanism) or integrating additional hardware or storage requirements either with clients or on APs.
- **Computational complexity:** This metric indicates whether the proposed defense mechanism incurs computational overhead in terms of processing, memory requirements. We use relative measures as follows: high (when servers, routers, or computers/laptops with comparatively high processing power or storage used), moderate (when PMF or any other additional authentication or verification mechanism used), and low (no extra resources or additional software used).
- **Technical overhead:** This metric indicates whether the proposed defense mechanism expects substantial technical knowledge on standard users to set up or operate. We use relative measures as follows: high (users have to install or set up new defense mechanisms on devices or install any proprietary software, update software/firmware, or any other sophisticated task), moderate (users have to configure or enable PMF on router or clients, and low (no task other than executing/running mechanisms).

Based on the above metrics, we evaluate stage 1 and 2 defense mechanism´s technical feasibility in Table 9.

**Table 9- Technical feasibility analysis of Multi-Channel MitM defense mechanisms**

| Metrics | Stage 1 defense mechanisms | | | | | Stage 2 defense mechanisms | | | | | | | |
|---|---|---|---|---|---|---|---|---|---|---|---|---|---|
| | (Vanhoef et al., 2018) | (Chatterjee et al., 2020) | (Vanhoef et al., 2020) | (Gong et al., 2020) | (Wi-Fi Alliance, 2020 § 6) | (Chin & Xiong, 2018) | (Naitik et al., 2018) | (Securingsam, 2017) | (Abare & Garba, 2019) | (Li et al., 2019) | (Cremers et al., 2020) | (Chothia & Ryan, 2020) | (SNORT, 2018) |
| Changes in the Wi-Fi standard | Required | Required | Required | Not Applicable | Not Applicable | Not Applicable | Not Applicable | Not Applicable | Required | Not Applicable | Required | Required | Not Applicable |
| Installation/ compatibility | Required on every device | Required on AP | Required on every device | Required on every client | Required on every device | Required on AP | Required on every client | Required on AP | Required on every device | Required on AP | Required on every device | Required on every device | Required on a client (pc/laptop) |
| PMF requirements | Required on every device | Not Applicable | Required on every device | Not Applicable | Required on every device | -----------This metric is not considered as it does not affect the implementation of stage 2 defense mechanisms-------------------- | | | | | | | |
| Firmware updates | Required on every device | Required on AP | Required on every device | Required on every client | Required on every device | Not Applicable | Not Applicable | Not Applicable | Required on every device | Not Applicable | Required on every device | Required on every device | Not Applicable |
| Third-party software /hardware integration | Not Applicable | Required on every AP (Additional hardware) | Not Applicable | Not Applicable | Not Applicable | Required on every AP (Additional hardware) | Not Applicable | Not Applicable | Not Applicable | Required on AP (Additional SDN software and storage) | Not Applicable | Not Applicable | Not Applicable |
| Computational complexity | Moderate | High | Moderate | Moderate | Moderate | High | High | Moderate | Moderate | High | Moderate | Moderate | High |
| Technical overhead | High | High | High | High | High | High | High | High | High | High | High | High | High |

#### 5.4.2. Discussion on evaluation of technical feasibility analysis

As seen in Table 9, we highlight that every mechanism incurs high technical overhead on common people in several ways. We give much importance to this because, ultimately, the defense mechanisms will be managed by people without much technical knowledge. The existing defense mechanisms may be effective theoretically, in laboratory settings, or in simulation environments; however, their practicality is quite difficult in IoT environments. This is mainly because:

1. Most defense mechanisms require Wi-Fi protocol standard changes that are hard to realize in practice, or the changes may take a long time for subsequent adoption by device vendors.
2. Almost every defense mechanism is required to install or configure their new solutions or specific network settings either on every client, AP, or both. This requirement considerably increases the technical burden on users. Besides, it is hard to achieve that all devices, especially IoT devices, will have to be modified, updated, or replaced by new defense mechanisms. Also, any unsupported device can still act as a vulnerable entity for MC-MitM enabled attacks.
3. Most stage 1 defense mechanisms depend entirely on PMF, but only some APs or router manufacturers support PMF. On the other hand, vendors rarely provide support for Wi-Fi clients. Enabling PMF might also require software/hardware updates on existing APs or clients. Additionally, PMF enforces advanced cipher suites or authentication mechanisms, which can be resource-intensive for IoT devices.
4. Firmware updating is a significant task that needs adequate technical knowledge. While implementing existing defense mechanisms, firmware updates are required in most cases as they mandate either changes in Wi-Fi standards or installing their new mechanisms. However, this requirement may be easy on robust devices but hard to achieve on every IoT device.
5. Integrating third-party software may be a difficult task in commercial or proprietary IoT environments as most of them may not always support it; moreover, such tasks are quite difficult for common people to set up themselves as the provision of technical support from IoT vendors is sometimes limited or non-existent. Besides, the said integration can increase the cost of maintenance, computational complexity, etc.

### 5.5. Summary

A significant concern stemming from the analysis of stage 1 defense mechanisms is the possibility of temporary MitM or insider MC-MitM attacks, especially with defense mechanisms included in WPA3. Although MC-defense mechanisms such as (Vanhoef et al., 2018) and (Vanhoef et al., 2020) are incorporated in 802.11 standards, they are not yet implemented in practice. On the other hand, most stage 2 defense mechanisms focus only on KRACK performed using the MC-MitM position. We could not find any stage 2 defense mechanisms in the literature related to MC-MitM enabled DoS attacks or the latest FragAttacks when writing this research paper. Similarly, the main takeaway from the feasibility analysis is that the existing defense mechanisms are not generalizable solutions to be practically implemented in IoT environments to effectively defend MC-MitM attacks. Further, we summarize the functionally, type of defense, advantages, and shortcomings of analyzed stage-1 and stage-2 defense mechanisms, respectively, in Tables 10 and 11.

## 6. Research Problems, Challenges, and Future Research Approaches

The state-of-the-art research analysis on MC-MitM attacks motivates us to highlight two kinds of research problems. These include: (i) **Design deficiencies** of the standard. (ii) **Technical infeasibility** issues of existing defense mechanisms, especially in Wi-Fi environments hosting IoT and outdated devices.

As **design deficiencies** of the standard, we glean the fact that there are no related works currently protecting PMF clients from MC-MitM attacks as they are able to circumvent and trouble PMF protection in many ways. This is a significant open research problem concerning both new WPA2 and WPA3 devices as they now mandate PMF. According to our analysis presented in Section 4.4, MC-MitM attacks also impact WPA3 networks in all modes of the operation. Additionally, MC-MitM attacks are especially critical if the attacks originate from insiders (e.g., fragmentation cache attack). This may result in the hijacking of private communication of other users or devices inside homes or offices. None of the existing defense mechanisms can effectively handle such a problematic situation. However, it is of great importance, and researchers can analyse the real impact of MC-MitM attacks on WPA3 devices or especially when WPA3 operates in its transition mode with several WPA2 devices. Most importantly, future defense mechanisms must consider protecting both PMF-capable and incapable devices, thereby protecting them from MC-MitM attacks.

Regarding the **technical infeasibility**, according to what we have analysed and summarized in Section 5, successful deployment of the existing MC-MitM attack defense mechanisms is hard in practice from an IoT perspective. This is an important open research problem that needs imperative developments against MC-MitM attacks in IoT environments like smart homes. To the best of our knowledge, there are no studies that analyse and propose IoT-friendly, hassle-free (without much user intervention and changes in existing devices) defense mechanisms for protecting IoT environments from MC-MitM attacks. Affordable and effective defense mechanisms must be developed because commercial IoT devices are deployed everywhere, in homes, buildings, and offices to stay connected. Nevertheless, it is important to safeguard such devices as they carry lots of sensitive information. MC-MitM attackers can trivially trick or hijack these IoT devices to loot sensitive information as most of them sometimes practice no encryption, low encryption strength, insufficient randomness, or weak key generation mechanisms, and can perform any other malicious or unintended activities.

Our study on MC-MitM attacks and state of the art detection systems also urges us to showcase the following essential research challenges, which shall be considered to have an improved defense against these attacks.

**Lack of sufficient backward compatibility** is one of the major concerns of existing MC-MitM defense mechanisms. As we highlighted in Section 4.3.1.1, most of the currently deployed WPA2 routers in our home or office settings still support WPA-TKIP through its transition mode. This is mainly to avoid interoperability issues and provide long-term support for outdated or constrained devices that sometimes support only TKIP. On the other hand, security patches, PMF, and new defense mechanisms are not practical on IoT networks with outdated or constrained devices. Consumers purchase several devices and expect to work longer, which means that such devices will be in Wi-Fi networks for several years while remaining as relatively weak entities in view of MC-MitM attacks. None of the existing detection systems have some practical backward compatibility considerations to safeguard old devices in our smart environments.

**Table 10- Summary of MC-MitM stage 1 defense mechanisms**

| Ref | Functionality | Type of defense | Advantages | Shortcomings |
|---|---|---|---|---|
| (Vanhoef et al., 2018) | Cryptographically authenticate or validate operating channels of AP and client during a 4-way handshake. | Prevention (Cryptographic method) | + Prevents channel misuse, so implicitly blocks multi-channel MITM attacks triggered by both base and improved variants.<br>+ Provides backward compatibility using Operating Channel Validation Capable (OCVC) flag in RSN fields.<br>+ Protects channel switch announcements (CSA) using PMF.<br>+ Incorporated in draft of 802.11 standard. | – Mandates change in WiFi standards.<br>– Mandates use of PMF which may not be always achievable.<br>– To take effect of OCI, both communicating parties must support it.<br>– Partial MITM is possible by blocking CSAs.<br>– Clients without OCI support remain vulnerable.<br>– Mandates firmware changes on Wi-Fi chips, which may be impractical for IoT devices. |
| (Chatterjee et al., 2020) | A PUF based challenge-response procedure to counteract the threat of fake access points. | Prevention (Cryptographic method) | + Prevents fake access points, so implicitly blocks multi-channel MITM attacks triggered by both base and improved variants.<br>+ Every client uniquely identifies every access point in a WLAN. | – Mandates change in WiFi standards<br>– PUF signatures (instances) of every AP must be created and stored in a separate server.<br>– Induces delay during 4-way handshake due to additional mutual authentication process.<br>– High technical overhead on users.<br>– Not suitable for commercial or proprietary IoT devices. |
| (Vanhoef et al., 2020) | Clients cryptographically authenticate beacons using an already distributed symmetric key from the AP. | Detection & Prevention (Cryptographic method) | + Prevents beacon spoofing. so implicitly blocks multi-channel MITM attacks triggered by both base and improved variants.<br>+ Detects and reports rogue AP.<br>+ Detects unauthenticated channel switch announcements (CSA)<br>+ Provides backward compatibility for older clients in identifying rogue APs.<br>+ Provides multiple BSSID beacon protection.<br>+ Incorporated in draft of 802.11 standard. | – Mandates change in WiFi standard<br>– Mandates use of PMF which may not be always achievable.<br>– Does not protect action frames and may not fully confront MC-MitM attacks.<br>– DoS attacks (flooding beacons) are inevitable.<br>– Beacon protection does not protect insider forgeries.<br>– Every client needs to store a reference beacon for verifying new beacons, which may be not ideal for IoT devices having constrained resources. |
| (Wi-Fi Alliance, 2020 § 6) | Clients identify the AP by verifying the digital signature of the AP´s public key and to block insider rogue APs | Prevention (Cryptographic method) | + Prevents insider rogue APs during the connection establishment.<br>+ Incorporated in WPA3 as an optional feature. | – MC-MitM attackers can bypass this method.<br>– Additional communication overhead due to digital signature verification.<br>– Useful only if every device supports this feature in WPA3. |
| (Gong et al., 2020) | Verify the anomalies in a pair of BSSID (MAC address of AP) and ESSID (Network name) when a client begins connecting to an AP. | Detection & Prevention (Anomaly detection Method) | + Detect rogue AP in a WLAN.<br>+ Alert the user.<br>+ Effective if the legitimate AP works on a specific channel. | – Requires changes in wpa_supplicant<br>– Every client in WLAN requires to install the modified wpa-supplicant.<br>– Ineffective if the AP operates on multiple channels, such as 2 GHz or 5 GHz.<br>– Detection rate becomes lower when continuous jamming is used.<br>– Integration may be difficult for proprietary IoT devices. |

**Table 11- Summary of MC-MitM stage 2 defense mechanisms**

| Ref | Functionality | Type of defense | Advantages | Shortcomings |
|---|---|---|---|---|
| (Chin & Xiong, 2018) | Detect privacy evasive attacks using KRACK scripts in a WLAN. | Detection | + Detect key reinstallation attacks on clients.<br>+ End-users get alerts without installing additional softwares. | – Unable to detect KRACK other than the attack on 4-way handshake.<br>– APs need integration of security modules.<br>– Increased computational (storage) and communication costs.<br>– High technical overhead on users.<br>– Not adoptable for IoT networks. |
| (Naitik et al., 2018) | Detect key reinstallation attacks by identifying duplicated EAPOL message 3 of the 4-way handshake in a target WLAN. | Detection | + Detect reuse of nonces during 4-way handshake on clients.<br>+ End-users get alerts if the system detects duplicate packets. | – Unable to detect KRACK other than the attack on 4-way handshake.<br>– Repeated handshake failures.<br>– Roaming issues.<br>– WPA key data can be forged.<br>– Verifying WPA key data in every frame incur huge computational costs.<br>– High technical overhead on users.<br>– Difficult to integrate into IoT environments. |
| (Abare & Garba, 2019) | Prevent KRACK attacks by encrypting entire messages in a 4-way handshake by using a new Boolean value to track key installation. | Prevention (Cryptographic method) | + Detect and mitigate reuse of nonces and key reinstallation attacks during 4-way handshake on clients. | – Unable to detect KRACK other than the attack on 4-way handshake.<br>– Mandates change in Wi-Fi standard.<br>– Handshake failure can happen even without the presence of an adversary.<br>– Increased computational costs due to additional calculation and verification of Boolean values during 4-way handshake.<br>– Probable delay in 4-way handshake. |
| (Li et al., 2019) | Defend key reinstallation attacks using SDN. | Detection & Prevention | + Detect and mitigate reuse of nonces and key reinstallation attacks during 4-way handshake on clients. | – Unable to detect KRACK other than the attack on 4-way handshake.<br>– APs needs integration of a SDN module.<br>– Increased computational (storage) and communication costs.<br>– Difficult to integrate into IoT environments. |
| (Cremers et al., 2020) | New formal models by properly using the nonces and replay counters of WPA2 handshake protocols. | Prevention (Cryptographic method) | + Detect key reinstallation attacks towards 4-way handshake, group key handshake, and WNM sleep mode.<br>+ Provides formal proof about the correctness of models. | – Requires additional security properties to be added to 802.11 standard.<br>– Conceptual models and not tested in real world attack settings. |
| (Chothia & Ryan, 2020) | New formal models with additional security properties against various KRACK attacks. | Prevention (Cryptographic method) | + Detect key reinstallation attacks towards 4-way handshake, group key handshake, and WNM sleep mode.<br>+ Provides formal proof about the correctness of models.<br>+ Detects security downgrade attacks (TKIP/CCMP). | – Requires additional security properties to be added to 802.11 standard.<br>– Conceptual models and not tested in real world attack settings. |
| (SNORT, 2018) | Identifies KRACK packets using SNORT rules. | Detection | + Detect any forms of key reinstallation attacks packets. | – Generate false alarms as contents used for KRACK packets can be found in other normal packets.<br>– Increased computational (storage) costs<br>– High technical overhead on users. |

**Rogue AP detection** as part of defending MC-MitM attacks can be challenging since the attacker cleverly spoofs almost every characteristic of the real AP and operates as legitimate in a WLAN (recall Section 3.6). As a result, such attackers can evade snooping-based rogue AP detection techniques by (Nikbakhsh et al., 2012). Usually, such detection strategies compare standard parameters of beacons, such as SSID, MAC addresses, channels, RSSI, sequence number, etc. However, the attacker can easily forge all these features if he knows them (Vanhoef et al., 2020). Communication channels can also be monitored or verified. But, blindly verifying the beacon's channel in a Wi-Fi medium may not be beneficial because there are many legitimate reasons for an AP to switch to different channels. Switching the channel is essential to avoid interference or noise in particular channels and is a dynamic action. Therefore, it may not be effective if we store an AP's channel to which a client was previously connected (Vanhoef & Piessens, 2014). Furthermore, since the attacker does not flood the network with beacons or probe requests, the frame arrival rate-based detection technique is also not helpful. Additionally, when the MC-MitM attacker uses special reactive jamming while establishing the MitM position, it would be hard to detect by IDS systems (Gong et al., 2020). In the above scenarios, it may be challenging to correctly distinguish MC-MitM attacks.

In light of the above research problems and challenges stemming from the analysis of MC-MitM attacks, we suggest that the best mitigation approach is a good intrusion detection strategy in line with the IoT environment's autonomous nature. It is worth researching on developing ideally generic, lightweight, and plug-and-play Intrusion Detection Systems (IDS). Such IDS solutions should easily integrate into any existing IoT infrastructure consisting of commercially available or proprietary IoT devices and easy to operate so that standard users without technical background can easily manage them. Solutions should also provide continuous security against different MC-MitM attack variants and generate instant alerts. To achieve this, researchers could study the fingerprints or signatures in terms of the patterns of attack variants (e.g., channel switch announcements sent through beacons, probe responses, action frames, malformed wireless frames appearance due to special kinds of channel jamming, frame exchanges etc.) and develop a signature-based IDS.

## 7. Conclusions

In this article, we have evaluated the capabilities of MC-MitM attacks and identified their distinct capabilities in manipulating protected Wi-Fi communications compared to traditional rogue AP MitM attacks. We have classified MC-MitM attacks, explored different kinds of related attacks in WPA, WPA2, WPA3, and analyzed their security impacts. Our analysis shows that MC-MitM attacks become more effective with the revelation of key reinstallation vulnerabilities, making such attackers decrypt communications from Wi-Fi devices unless appropriately patched. Though some patches are available, they do not apply to every Wi-Fi device. In this regard, we have identified significant security patching difficulties, especially on IoT devices. With the entry of recent FragAttacks, MC-MitM attacks become more widespread and practical to inject genuine packets into protected wireless networks and obtain user´s sensitive data. FragAttacks again brought huge challenges and a matter of security concern. Devices are likely vulnerable in the coming years due to the lack of proper implementation of Wi-Fi Alliance patches and adequate defense mechanisms. We can expect the same difficulties of KRACK patching with FragAttacks.

We identified that PMF could not be an adequate deterrent as it can be easily circumvented through MC-MitM attacks. Our studies shed light on the fact that MC-MitM attacks impact WPA3 networks in several ways due to their ability to circumvent PMF protection. We highlight this is a significant problem because WPA3 networks are evolving in our home and office environments. As far as MC-MitM defense is concerned, on the one hand, the existing mechanisms are not adequate as most of them allow some forms of insider MC-MitM attacks. On the other hand, MC-MitM attack defense remains an open research problem, especially from an IoT´s perspective. We presented a technical feasibility analysis of the existing defense mechanisms, which uncovered that they are not flexible to be deployed in proprietary IoT networks consisting of constrained Wi-Fi-based IoT sensors and controllers.

This article gives the research community a view of MC-MitM attacks, their characteristics, and a fundamental understanding of the inner workings of various MC-MitM enabled attacks. It also highlights the importance of protecting Wi-Fi and IoT networks, especially when connected devices are working on different Wi-Fi protected access protocols and existing mechanisms cannot be practiced. To this end, we suggest developing lightweight and effective wireless intrusion detection systems for particularly defending against MC-MitM attacks in real Wi-Fi based IoT networks.

### CRediT authorship contribution statement

**Manesh Thankappan**: Conceptualization, Methodology, Investigation, Writing - original draft, Visualization, Software.
**Helena Rifà-Pous**: Conceptualization, Validation, Writing - review & editing, Supervision, Resources, Funding acquisition.
**Carles Garrigues**: Conceptualization, Validation, Writing - review & editing, Supervision, Resources, Funding acquisition.

### Acknowledgements

This work was partly funded by the Spanish Government through project **RTI2018-095094-B-C22** “**CONSENT**”. We would also thank Mathy Vanhoef for his help by answering some of our queries and providing source codes related to MC-MitM attacks.